\documentclass[journal]{IEEEtran}
\usepackage[caption=false,font=normalsize,labelfont=sf,textfont=sf]{subfig}
\usepackage{cite}
\usepackage{amsmath}		
\usepackage{amsfonts}		
\usepackage{amssymb}        
\usepackage{ulem}
\ifCLASSINFOpdf
\usepackage{graphicx}
\usepackage{txfonts}
\else
\usepackage{graphicx}
\fi
\usepackage{float}
\usepackage{multicol}
\usepackage{epstopdf}
\usepackage{breqn}
\usepackage{multirow}
\usepackage{caption}
\usepackage{soul}
\usepackage{algorithm}
\usepackage{algpseudocode}
\usepackage{mathtools}
\usepackage{xcolor}

\usepackage{bbm}
\usepackage{dblfloatfix}

\usepackage{hyperref}
\hypersetup{
    colorlinks=true,
    linkcolor=blue,
    filecolor=magenta,      
    urlcolor=blue,
}
    
 \usepackage{bm}

\newcommand{\ma}  [1]{ \bm{#1} } 
\newcommand{\IndexV}[2]{ #1\left[#2\right] } 

\begin{document}

\title{Deep Neural Network Augmented Wireless Channel Estimation for {\color{black} Preamble-based OFDM PHY on Zynq} System on Chip}
\author{Syed Asrar ul haq, Abdul Karim Gizzini, Shakti Shrey, Sumit J. Darak, Sneh Saurabh and Marwa Chafii
\thanks{This work is supported by the funding received from core research grant (CRG) awarded to Dr. Sumit J. Darak from DST-SERB, GoI.

Syed Asrar ul haq, Shakit Shrey, Sumit J. Darak and Sneh Saurabh are with Electronics and Communications Department, IIIT-Delhi, India-110020 (e-mail: \{syedh,shakti20323,sumit,sneh\}@iiitd.ac.in)

Abdul Karim Gizzini is with ETIS, UMR8051, CY Cergy Paris Université, ENSEA, CNRS, France (e-mail: abdulkarim.gizzini@ensea.fr).

Marwa Chafii is with the Engineering Division, New York University (NYU) Abu Dhabi, 129188, UAE, and NYU WIRELESS, NYU Tandon School of Engineering, Brooklyn, 11201, NY (e-mail: marwa.chafii@nyu.edu).
}}

\maketitle

\begin{abstract}
Reliable and fast channel estimation is crucial for next-generation wireless networks supporting a wide range of vehicular and low-latency services. Recently, deep learning (DL) based channel estimation has been explored as an efficient alternative to conventional least-square (LS) and linear minimum mean square error (LMMSE) approaches. Most of these DL approaches have not been realized on system-on-chip (SoC), and preliminary study shows that their complexity exceeds the complexity of the entire physical layer (PHY). The high latency of DL is another concern. This paper considers the design and implementation of deep neural network (DNN) augmented LS-based channel estimation (LSDNN) for { \color{black} preamble-based orthogonal frequency-division multiplexing (OFDM) physical layer (PHY)} on SoC. We demonstrate the gain in performance compared to the conventional LS and LMMSE approaches. Via software-hardware co-design, word-length optimization, and reconfigurable architectures, we demonstrate the superiority of the LSDNN over the LS and LMMSE for a wide range of signal-to-noise ratio (SNR), number of pilots, preamble types, and wireless channels. Further, we evaluate the performance, power, and area (PPA) of the LS and LSDNN application-specific integrated circuit (ASIC) implementations in 45 nm technology. We demonstrate that word-length optimization can substantially improve PPA for the proposed architecture in ASIC implementations.

\end{abstract}

\section{Introduction} \label{introduction}


Accurate channel state information estimation is crucial for reliable wireless networks \cite{Zeadally2020ITSstandard, Mumtaz2021lowLatency6g}.
Conventional least square (LS) based channel estimation is widely used in long-term evolution (LTE) and WiFi networks due to its low complexity and latency. However, it performs poorly at a low signal-to-noise ratio (SNR) \cite{6814271}. Other techniques like linear minimum mean square error (LMMSE) offer better performance than LS but need prior knowledge of noise, and second-order channel statistics \cite{7447802}. Furthermore, it is computationally intensive \cite{7447802,van1995channel}. 
The limited sub-6 GHz spectrum has led to the co-existence of heterogeneous devices with limited guard bands. This has resulted in poor SNR due to high noise floor and adjacent-channel interference. This has led to significant interest in improving the performance of the conventional LS and LMMSE channel estimation approaches. With the integration of vehicular and wireless networks, the radio environment has become increasingly dynamic, posing serious challenges in establishing reliable communication links. Furthermore, high-speed ultra-reliable services demand ultra-low latency of the order of milliseconds \cite{9247524}.

Recent advances in artificial intelligence, machine and deep learning (AI-MDL) have been explored to improve the performance of wireless physical layer such as channel estimation, symbol recovery, link adaptation, localization, etc. \cite{DLPHY2,yang2019deep,wei2020deep,ma2020data,gizzini2020enhancing,gizzini2020deep,pan2021channel,DLPHY1,DL_Rohith}.  Various studies have shown that data-driven AI-MDL approaches have potential to significantly improve the performance and offer increased robustness towards imperfections, non-linearity, and dynamic nature of the wireless environment and radio front-end \cite{8839651, 8233654, 9508930, 9247524,9440456}. In the June 2021 3GPP workshop, various industry leaders presented the framework for the evolution of intelligent and reconfigurable wireless physical layer (PHY), emphasizing the commercial potential of the AI-MDL based academic research in the wireless domain. Though intelligent and reconfigurable PHY may take a few years to get accepted commercially, some of these initiatives are expected to be included in the upcoming 3GPP Rel-18.

 
Few works have replaced the LS, and LMMSE channel estimation with computationally complex deep learning (DL) architectures \cite{gizzini2020enhancing,8640815,8944280}. However, most existing works have not been realized on system-on-chip (SoC). {\color{black} In this context, this work focuses on improving the channel estimation performance in preamble-based orthogonal frequency division multiplexing (OFDM) wireless PHY used in IEEE 802.11p and similar standards.}
Compared to existing works, we augment the LS with \textcolor{black}{a Fully Connected Feedforward} DNN instead of replacing it with computationally intensive DL networks. We demonstrate the performance gain over the conventional LS and LMMSE approaches. We extensively study the effects of training SNR on the performance of the LSDNN channel estimation.

{\color{black} We have mapped the proposed and existing algorithms on two Zynq multi-processor system-on-chip (ZMPSoC) boards from AMD-Xilinx: ZC706 and ZCU111. The ZC706 comprises a dual-core ARM Cortex-A9 processor and Xilinx 7 series Kintex FPGA. The ZCU111 comprises a quad-core ARM Cortex A53 processor and an ultra-scale FPGA.} Through software-hardware co-design, word-length optimization, and {\color{black} adaptable} architectures, we demonstrate the superiority of the proposed LSDNN architecture over LS and LMMSE architectures for a wide range of SNRs, number of pilots, preamble types, and wireless channels. Experimental results show that the proposed LSDNN architecture offers lower complexity and a huge improvement in latency over the LMMSE. Next, we evaluate the performance, power, and area (PPA) of the LS and LSDNN on application-specific integrated circuit (ASIC) implementations in 45 nm technology. We demonstrate that the word-length optimization can substantially improve PPA for the proposed architecture in ASIC implementations. The AXI-compatible hardware IPs and PYNQ-based graphical user interface (GUI) demonstrating the real-time performance comparison on the ZMPSoC are physical deliverables of this work. {\color{black} Please refer to \cite{code} for source codes, datasets, and hardware files used in this work.} The work presented in this paper is a significant extension of the conference paper \cite{gizzini2020enhancing}. Here, we improve the LSDNN algorithm as per the memory constraints of the SoC, two LSDNN models, and detailed performance analysis, including the effect of training SNR. The hardware architectures on ZMPSoC (Section~~\ref{sec:hwImpl} onwards) have not been discussed previously.

The rest of the paper is organized as follows. Section~\ref{sec:litRev} reviews the work related to applications of AI-MDL approaches for wireless PHY. In Section~\ref{sec:sysMod}, we present the system model and introduce LS, and LMMSE-based channel estimation approaches. The proposed LSDNN algorithm and simulation results comparing the performance of LS, LMMSE, and LSDNN are given in Section~\ref{sec:DNN} followed by their architectures in Section~\ref{sec:hwImpl}. The functional performance and complexity results of fixed-point architectures are analyzed in Section~\ref{sec:results}. The ASIC implementation and results are presented in Section~\ref{sec:ASIC Implementation}. Section~\ref{sec:conclusions}
concludes the paper along with directions for future works.

\textbf{Notations}: Throughout the paper, vectors are defined
with lowercase bold symbols $\ma{x}$ whose $k$-th element is $\ma{x}[k]$. Matrices are written as uppercase bold symbols $\ma{X}$. 
\section{Literature Review: DL in Wireless PHY}\label{sec:litRev}


Various studies and experiments have demonstrated that conventional frameworks such as Shannon theory \cite{shannon1948mathematical}, detection theory \cite{kay1998fundamentals}, and queuing theory \cite{karray2013queueing} based wireless PHY suffer from performance degradation due to randomness and diversity of wireless environments. Instead of exploring approximate frameworks, recent advances in AI-MDL approaches and their ability to address hard-to-model problems offer a good alternative for making the wireless PHY robust \cite{singh2021toward,darak2019multi}. Numerous works have shown that traditional ML techniques have been successful in combating complex wireless environments and hardware non-linearities \cite{CESoC4,CESoC5}. However, they need manual feature selection. The DL approaches offer the capability to extract features from the data itself, thereby eliminating the need for manual feature extraction. This has lead to numerous AI-MDL based approaches for various problems such as modulation classification \cite{zhou2020deep,9174643},  
spectrum sensing \cite{Himani_DL,DL_Rohith}, multiple-input multiple-output (MIMO) high-resolution channel feedback \cite{wang2018deep},
mobility and beam management \cite{zhou2019Beammanagement}, link adaptation \cite{5200378}.
With the evolution of heterogeneous large-scale networks, existing frameworks suffer from scaling issues due to the high cost of exhaustive searching and iterative heuristic algorithms \cite{shi2011iteratively}.
Such scaling issues can be handled more efficiently using AI-MDL approaches \cite{zhang2021scalable}. 
 To bring DL-aided intelligent and reconfigurable wireless PHY into reality, issues such as potential use cases, achievable gain and complexity trade-off, evaluation methodologies, dataset availability, and standard compatibility impact need to be addressed \cite{8839651,  9508930, 9247524,9440456}.

The DL-based PHY can be categorized in two approaches: 1)~Complete transmitter and receiver are replaced with respective DL architectures \cite{8054694,dorner2018Endtoend}, 2)~Independent DL blocks for each sub-block or combination of sub-blocks in PHY \cite{8052521,9174643,8640815,9739166,8944280}. The drawback of the first approach is that it may not provide intermediate outputs such as channel status indicator, precoding, and channel rank feedback, making them incompatible with existing standards. Furthermore, existing works can not support high throughput requirements. In this paper, we focus on the second approach addressing the channel estimation task in wireless PHY.

ChannelNet \cite{8640815} treats the LS estimated values at pilot sub-carriers in an OFDM frame as a low-resolution 2D image and develops a CNN-based image enhancement technique to increase the resolution to obtain accurate channel estimates at data sub-carriers. Its performance is sensitive to the SNR conditions, which demand SNR-specific models resulting in high reconfigurable time and large on-chip memory. This model is further enhanced \cite{8944280} by replacing the two CNN models with a single deep residual neural network (ReEsNet) to reduce its complexity and improve performance. Interpolation-ResNet in \cite{9739166} replaces the transposed convolution layer of ReEsNet with a bilinear interpolation block to reduce the complexity further and make the network flexible for any pilot pattern. The computational complexity and memory requirements of these models are high, even higher than the complexity of PHY. In this work, we focus on augmenting the conventional channel estimation with DL instead of DL-based channel estimation \cite{8640815,8944280,9739166}.


The validation on synthetic datasets is another concern in DL-based approaches. This has been addressed in some of the recent works, such as \cite{9250028} where DL-based channel estimation is performed using real radio signals, demonstrating the feasibility in practical deployment. This works provides a detailed process and addresses the concerns regarding dataset creation from the RF environment and training and testing the models using real radio signals. The authors in \cite{dorner2018Endtoend} used the concept of transfer learning in \cite{pan2010TransferLearning} to devise a two-phase training strategy to train a DL-based PHY using real-world signals and demonstrated over-the-air communication. It offered similar performance as conventional PHY validating the feasibility of DL-based approaches in a real-radio environment.
However, limited efforts have been made on hardware realization of the DL-based wireless PHY \cite{CESoC1,CESoC2,CESoC3,ACM_DLCESD,Sai_tnnls}. Hence, there is limited knowledge of their performance on fixed-point hardware, latency, and computation complexity. The impact of the hardware-specific constraints such as high cost and area of on-chip memory and a limited number of memory ports on the training approaches have not been highlighted in the literature yet. The work presented in this paper addresses these issues and offers innovative solutions at algorithm and architecture levels for channel estimation in wireless PHY.



\section{System Model}\label{sec:sysMod}

\label{Sec:Systemmodel}

\begin{figure}[t]
\centering
\includegraphics[scale=0.65]{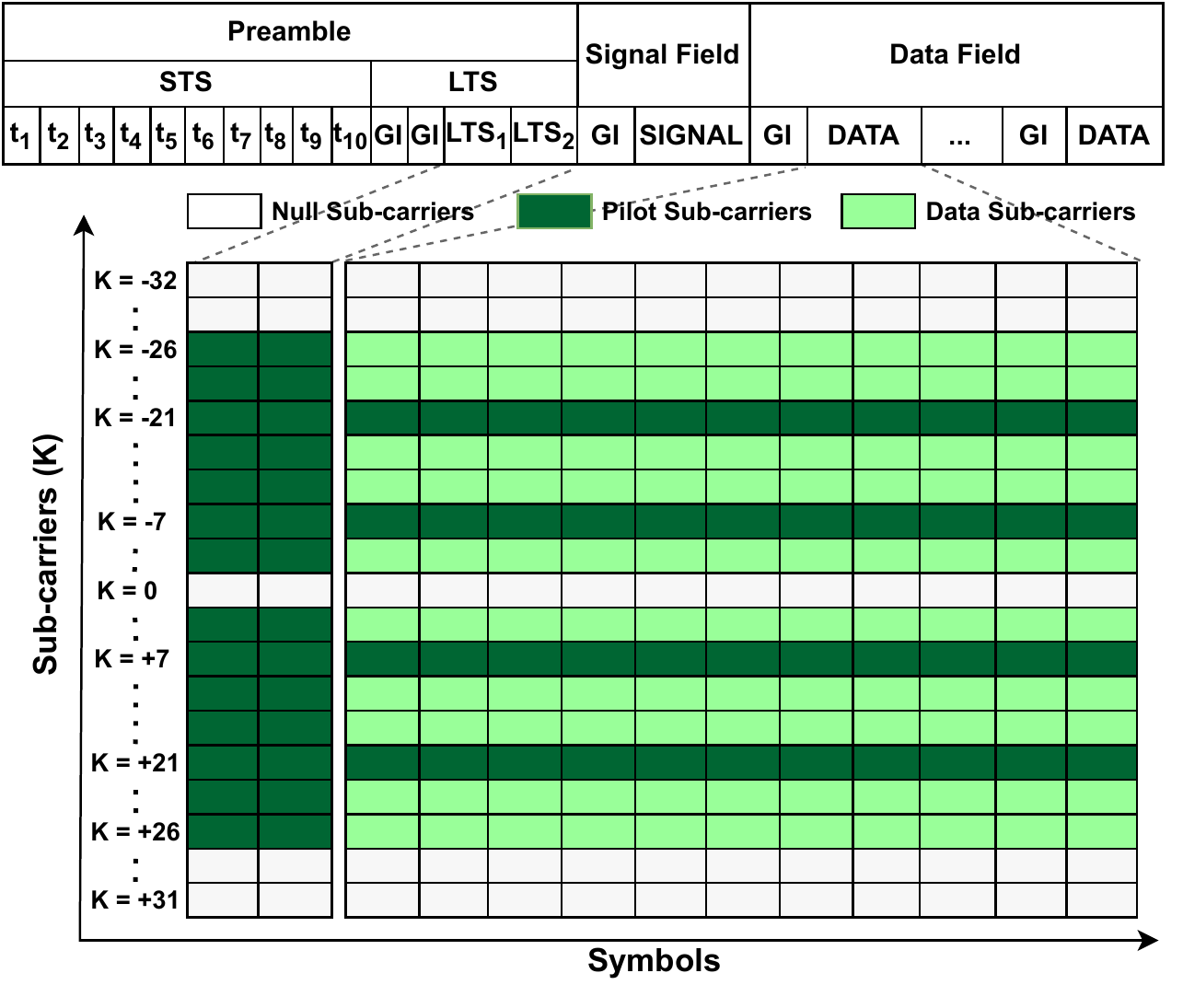}
\caption{\small  IEEE 802.11p frame structure and sub-carrier mapping.}
    \label{fig:frameStruct}
\end{figure}

This paper considers orthogonal frequency-division multiplexing (OFDM) based transceivers with frame structure based on IEEE 802.11p standard shown in Fig. \ref{fig:frameStruct}.
{\color{black} The transmitted frame consists of a preamble header including ten short training sequences and two long training symbols (LTS), followed by the signal and data fields. The LTS consists of predefined fixed symbols known to both transmitter and receiver and used for channel estimation at the beginning of the frame.}
There are total $K = 64$ sub-carriers with the sub-carrier spacing of $156.25$ KHz, and the corresponding transmission bandwidth is $10$ MHz. Each LTS consists of $64$ sub-carriers in a single OFDM symbol, of which $12$ are NULL sub-carriers, and $52$ are sub-carriers carrying BPSK modulated preamble sequence. Each data symbol carries $48$ data and $4$ pilot sub-carriers. In this work, we assume that the channel is static over the course of frame transmission. {\color{black}In cellular PHY, the preamble is absent, and comb-pattern-based pilots are embedded in OFDM data symbols \cite{8052521,9174643,8640815,9739166,8944280,ACM_DLCESD}. Due to different system models and channel estimation approaches, we limit our discussion to preamble-based channel estimation.}

\begin{figure}[!t]
\centering
\includegraphics[scale=0.53]{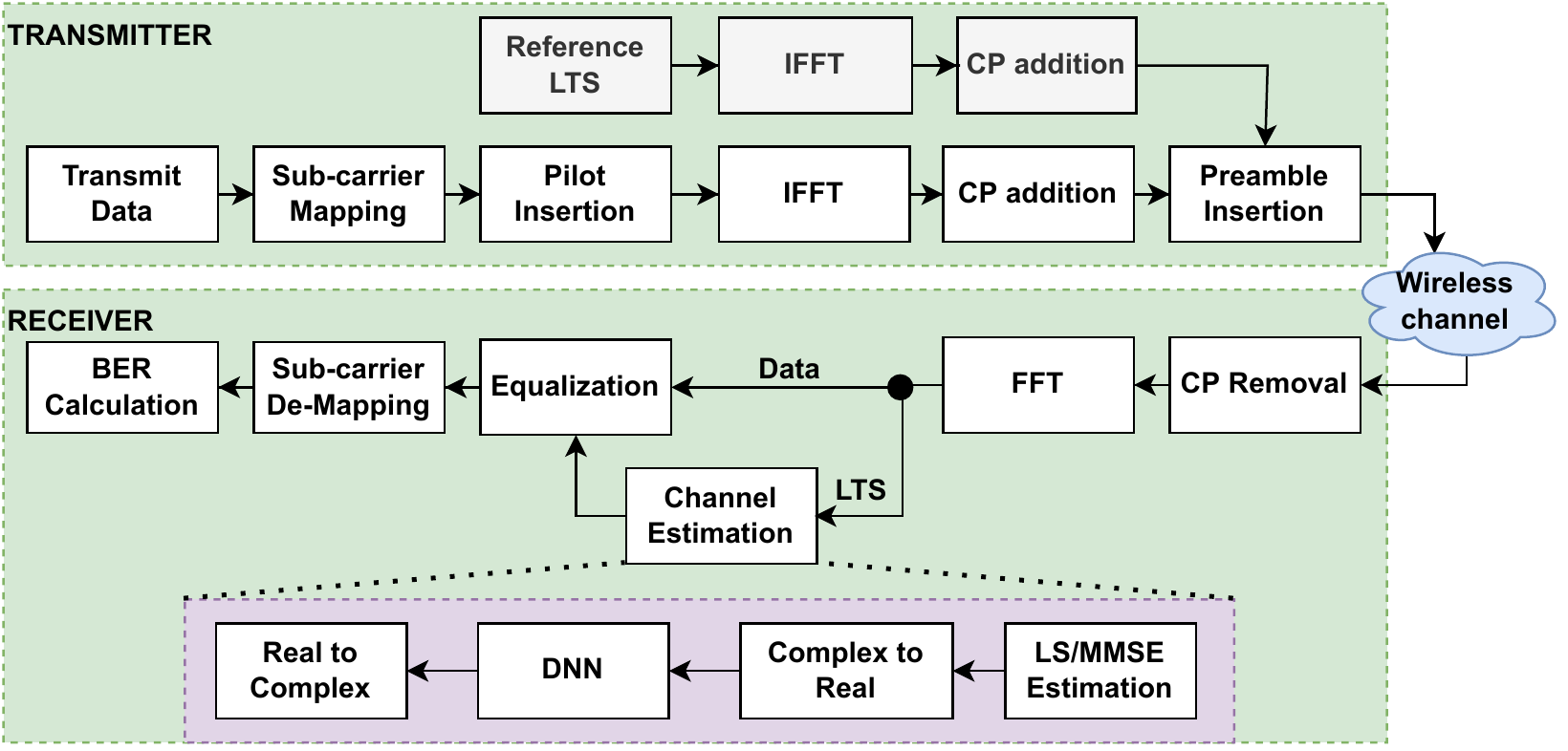}
\caption{\small  Building blocks of the IEEE 802.11p based OFDM transmitter and receiver PHY.}
    \label{fig:sysDesign}
\end{figure}

The block diagram of the IEEE 802.11p  PHY transceiver is shown in Fig.~\ref{fig:sysDesign}. The data to be transmitted is modulated using an appropriate modulation scheme such as QPSK/QAM, followed by sub-carrier mapping. The set of 64 sub-carriers is passed through OFDM waveform modulation comprising IFFT and cyclic prefix addition. The length of cyclic prefix is $K_{cp} = 16$. The scheduler inserts the preamble OFDM symbols at the beginning of each frame, comprising a certain number of data symbols. Next, the complete OFDM frame is transmitted through the wireless channel.
At the receiver, OFDM symbols containing the LTS preamble are detected and extracted, followed by OFDM demodulation. Then, channel estimation and equalization operations are performed.



The frequency-domain input-output relation between the transmitted and the received OFDM frame at the receiver can be expressed as follows

\begin{equation}\label{eqn:2}
    \ma{Y}[k,i] = {\ma{H}}[k,i] \ma{X}[k,i] + \ma{V}[k,i].
\end{equation}

Here, $\ma{X}[k,i]$,  ${\ma{H}}[k,i]$,  $\ma{Y}[k,i]$, and  $\ma{V}[k,i]$ denote the transmitted OFDM frame, the frequency domain channel gain, the received OFDM frame, and the addictive white Gaussian noise (AWGN) with zero mean and variance, $\mathcal{N}_0$. $k$ and $i$ denote the sub-carrier and OFDM symbol indices within the received frame, respectively. 

Recall that the channel estimation is performed once per frame using the received preamble symbols, therefore~\eqref{eqn:2} can be rewritten as
\begin{equation}
\ma{Y}[k, p] = {\ma{H}}[k,p] \ma{D}[k,p] + \ma{V}[k,p],
\label{in-op}
\end{equation}
where $\ma{D}[k,p] = \IndexV{\ma{d}_p}{k}$ denote the $p$-th predefined transmitted preamble symbol, where $1 \le p \le K_p$ and $K_{p}$ is the total number of the transmitted preamble symbols within the frame.




In this context, the LS channel estimation~\cite{7447802} can be expressed as
\begin{equation}
     \hat{\tilde{\ma{H}}}_{\text{LS}}[k,p] = \frac{\sum_{q=1}^{K_p} \ma{Y}[k,q]}{K_p \ma{D}[k,p]}
\label{eq:LS}     
\end{equation}


The LS estimation is easy to implement in hardware and does not require prior channel and noise information. 
On the other hand, the LMMSE channel estimation~\cite{7447802} needs prior knowledge of the second order channel and noise statistics. Mathematically, the estimated channel gain at the $k$-th sub-carrier can be expressed as 
\begin{equation}\label{eqn:5}
    \hat{\tilde{\ma{H}}}_{\text{LMMSE}}[k,p] = {\ma{R}_h}_{p}\left ({\ma{R}_h}_{p} +\left ( \frac{K\mathcal{N}_0}{E_p} \right ) \ma{I}\right )^{-1}\hat{\tilde{\ma{H}}}_{\text{LS}}[k,p],
\end{equation}
where ${R_h}_p$ is the channel auto-correlation matrix, and ${E}_{p}$ denotes the power per preamble symbol. It can be observed that the LMMSE improves the performance of the LS, and the improvement depends on the prior knowledge of the channel and noise statistics.




\section{Deep Neural Networks Based Channel Estimation}\label{sec:DNN}

In this section, we present LSDNN design details and simulation results using floating-point arithmetic comparing various channel estimation approaches. 

\subsection{LSDNN based Channel Estimation}

\begin{figure} [!b]
\centering
\includegraphics[scale=0.56]{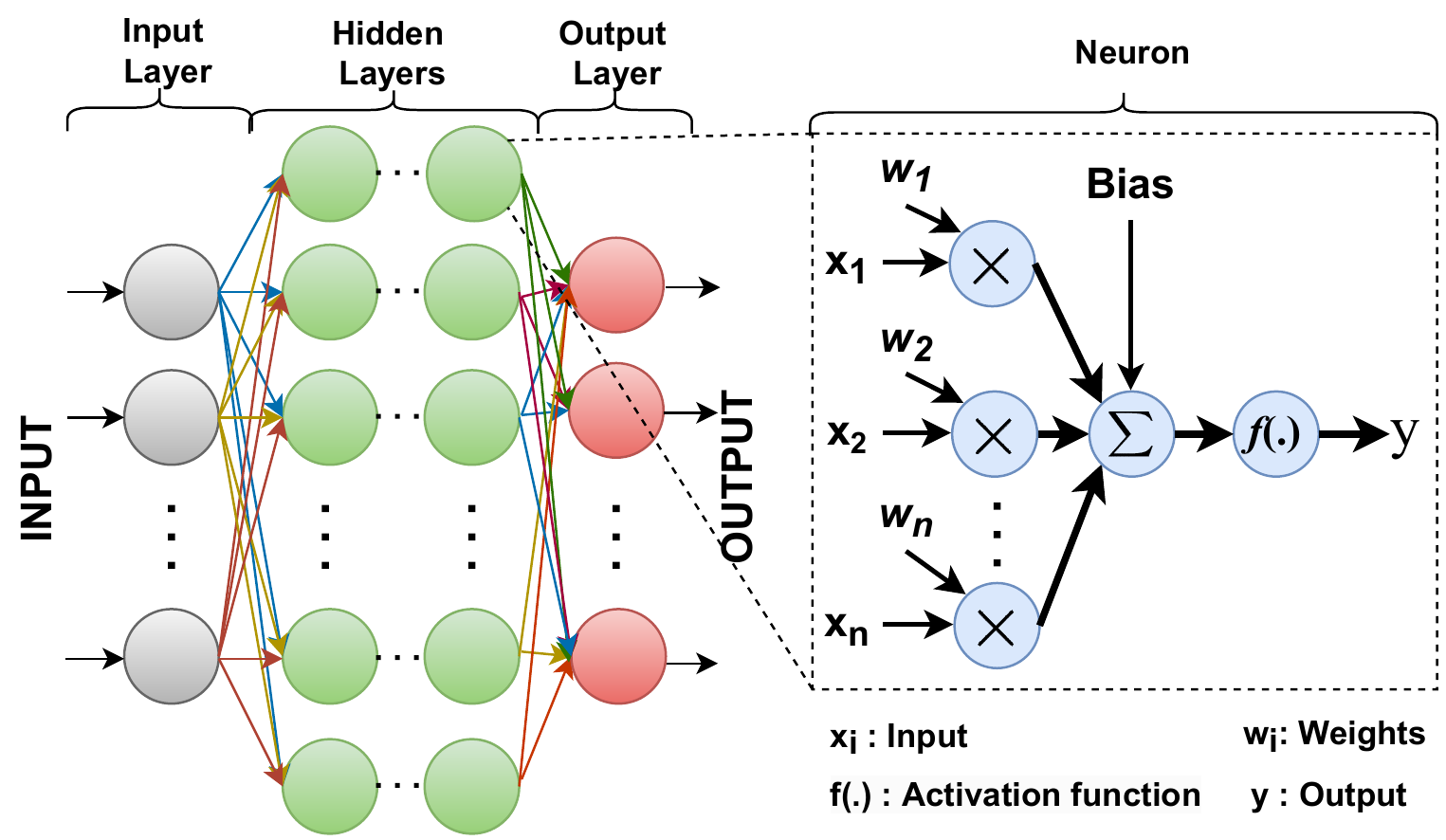}
\caption{\small  Fully connected neural network.} 
    \label{fig:neural}
\end{figure}

 \begin{figure*}[!b]
        \centering
        \vspace{-0.25cm}
        \includegraphics[scale=0.58]{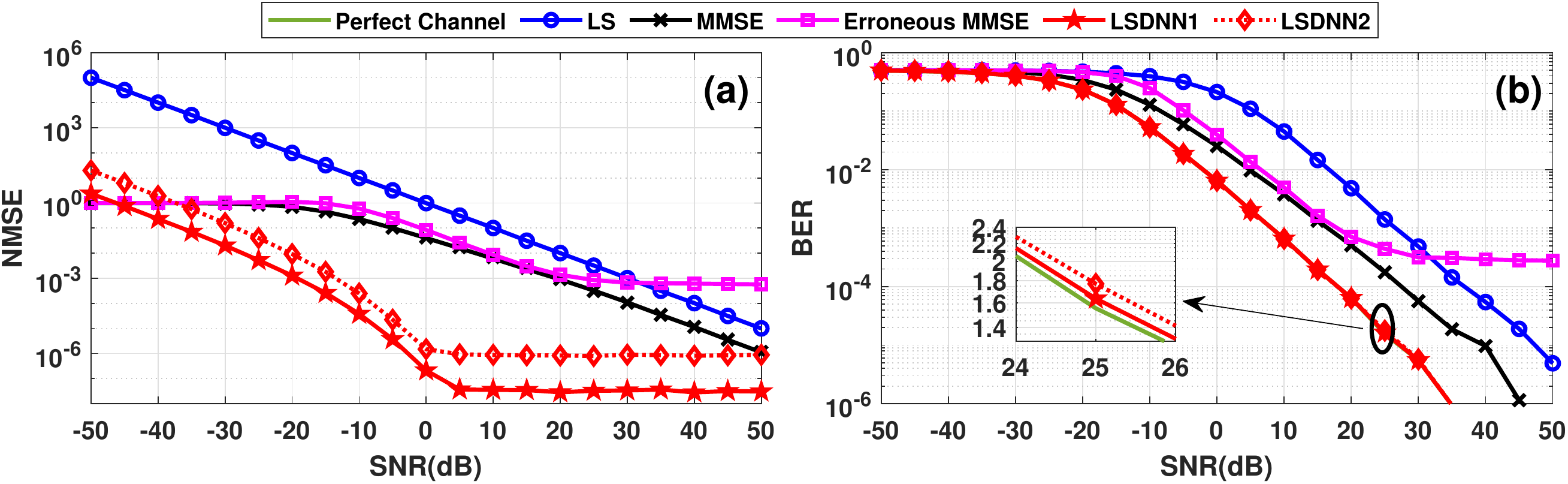}
         \vspace{-0.1cm}
      \caption{(a) NMSE and (b) BER comparisons of various channel estimation approaches using floating point arithmetic.}
\label{fig:comparison}
        \vspace{-0.1cm}
\end{figure*}

Deep Neural Networks (DNN) is a subset of DL that tries to mimic how information is passed among biological neurons. A fully-connected neural network is shown in Fig. \ref{fig:neural}. It has a layered structure, with each layer feeding to the next layer in a forward direction. The first layer is called an input layer representing the input to the DNN, and the last layer is called an output layer from which the outputs are taken. The layers between the input layer and output layer are called hidden layers. The output and hidden layers contain parallel processing elements called neurons, which take the inputs from the previous layer and send output to the next layer. A neuron performs the summation of weighted inputs followed by a non-linear activation function, as shown in Fig.\ref{fig:neural}. These functions include Sigmoid, tanh, ReLU, leaky ReLU \cite{sze2017efficient}. For DNN with $L$ layers, including input and output layers, the output of $j$-th neuron in layer $l$, $1 < l < L$ is given as

\begin{equation}\label{eqn:6}
   \ma{y}_j^l = f^{(l)}\left ( \sum_{i=0}^{N_l}\left ( \ma{w}_{i,j}^l \ma{y}_{i,j}^{l-1} \right )+ \ma{b}_j^l \right ),
\end{equation}
where $N_l$ is the total number of neurons in $l$-th layer, $b_j^l$ is the bias of $j$-th neuron of $l$-th layer, $w_{i,j}$ is the weight from $i$-th neuron of $(l-1)$-th layer to $j$-th neuron of $l$-th layer, and $f^{(l)}\left ( .\right )$ is the activation function of $l$-th layer.

{\color{black} The proposed DNN augmented LS estimation approach, an extension of \cite{gizzini2020enhancing}, processes the output of the LS estimator using the DNN, thereby reducing the effect of noise on LS estimates. As shown in Fig.~\ref{fig:sysDesign}, the proposed LSDNN channel estimator proceeds as follows

{\color{black}

\begin{itemize}
    \item LS channel estimation as illustrated in~\eqref{eq:LS}, where $\hat{\tilde{\ma{H}}}_{\text{LS}}  \in \mathbb{R}^{K_{\text{on} \times 1}}$ and $K_{\text{on}}$ refers to the active subcarriers within the transmitted preamble symbols.

    \item $\hat{\tilde{\ma{H}}}_{\text{LS}}  \in \mathbb{R}^{K_{\text{on} \times 1}}$ is converted from complex to real-valued domain by stacking the real and imaginary values in one vector such that $\hat{\tilde{\ma{H}}}^{(R)}_{\text{LS}}  \in \mathbb{R}^{2K_{\text{on} \times 1}}$.

     \item $\hat{\tilde{\ma{H}}}^{(R)}_{\text{LS}}  \in \mathbb{R}^{2K_{\text{on} \times 1}}$ is fed as an input to the employed DNN.

     \item Finally, the processed LS complex-valued channel estimates $\hat{\tilde{\ma{H}}}^{(DNN)}_{\text{LS}}  \in \mathbb{R}^{K_{\text{on} \times 1}}$ is obtained by converting the real-valued output vector of DNN back to the complex-valued vector.
\end{itemize}

A training dataset of 30,000 samples is generated using MATLAB-based IEEE802.11p PHY. The transmitter transmits the LTS over the channel for a wide range of SNR. The LS estimate of received LTS and the corresponding channel frequency response are used as the training data. The dataset is then divided into a training set of 24,000 samples and a validation set of 6,000 samples in an 80:20 ratio. During inference, the OFDM frame consisting of LTS and data symbols is transmitted over the channel. The LS estimate of LTS is passed as input to the trained DNN, and the estimates thus obtained are used to equalize the data symbols for subsequent BER analysis. Please refer to \cite{code} for source codes, datasets, and hardware files used in this work.}

The DNN is trained to reduce the loss function $J_{\Omega ,b}(\tilde{H} , y_{\hat{\tilde{\ma{H}}}_{\text{LS}}}^{L})$ over one long training symbol, where $\ma{H}$ is the original channel frequency response and $\hat{\tilde{\ma{H}}}_{\text{LS}}$ is the channel estimated using the LS approach. Training is an iterative process where the model parameters are updated using optimization functions to minimize the loss over time. Various optimization functions such as stochastic gradient descent, root mean square prop, and adaptive moment estimation (ADAM) can be used \cite{ruder2016overview}.

Hyper-parameters for the DNN architecture is chosen manually. Multiple iterations were performed with different hyper-parameters until the one with the best accuracy and low latency was found. We restricted our search to only a few hidden layers to reduce the complexity of the overall design. For illustrations, we consider two DNN architectures: 1) LSDNN1: DNN with a single hidden layer of size $K_{on}$, and 2) LSDNN2: DNN with two hidden layers of size $2\times K_{on}$. Each model is trained for 500 epochs with Mean Square Error (MSE) as a loss function and ADAM as an optimizer. Table \ref{tab:table1} summarises the specifications of neural network.  After training, the next step is inference, in which a trained DNN model is tested on new data to evaluate its performance. The hardware realizations of DNN focus on accelerating the inference phase.}



\begin{table} [!b]
\centering
\caption{\small DNN Parameters}
\label{tab:table1}
\renewcommand{\arraystretch}{1.1}
 \resizebox{\columnwidth}{!}{
\begin{tabular}{|l|l|}
\hline
\textbf{Parameters}                        & \textbf{Values}         \\ \hline
LSDNN1 (hidden layers ; neurons per layer) & (1 ; $K_{on}$)          \\ \hline
LSDNN2 (hidden layers ; neurons per layer) & (2; $2 \times K_{on}$ ) \\ \hline
Activation Function                        & ReLU                    \\ \hline
Epochs                                     & 500                     \\ \hline
Optimizer                                  & ADAM                    \\ \hline
Loss Function                              & MSE                     \\ \hline
\end{tabular}}
\end{table}

 \begin{figure*}[!t]
        \centering
        \vspace{-0.2cm}
        \includegraphics[scale=0.575]{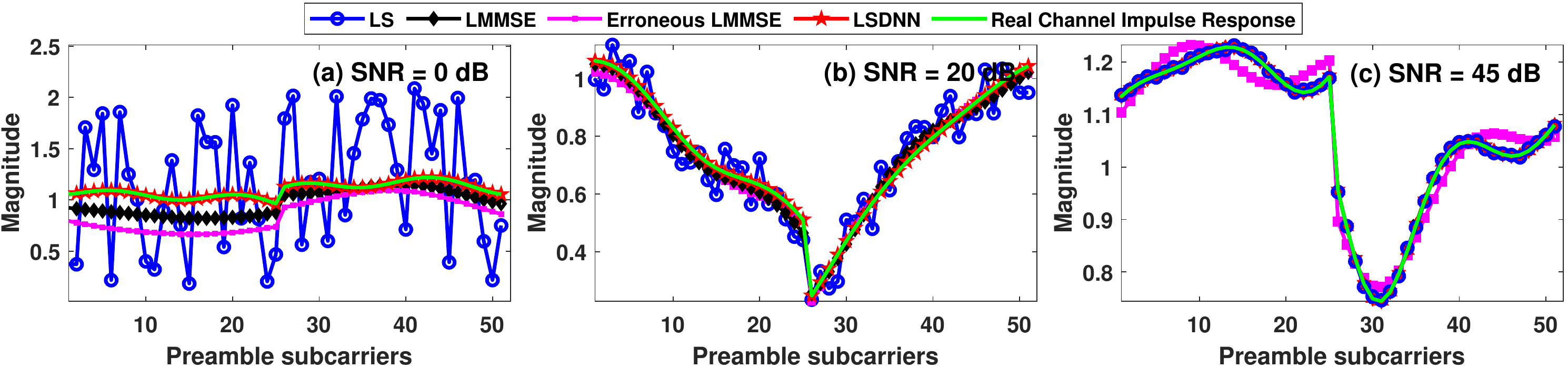}
         \vspace{-0.1cm}
     \caption{Magnitude plot for channel estimation corresponding to subcarriers in an OFDM symbol for (a) 0dB SNR, (b) 20dB SNR, (c) 45dB SNR.}
\label{fig:mag}
        \vspace{-0.2cm}
\end{figure*}
\subsection{Simulation Results on Floating Point Arithmetic}
\label{Sec:SimRes}


For the system model discussed in Section~\ref{sec:sysMod}, we have used six channel models considering vehicle-to-vehicle (V2V) and roadside-to-vehicle (RTV) environments as discussed in \cite{Acosta2007ChModels}. We have considered two performance metrics: 1) Normalized mean square error (NMSE) and 2) Bit-error-rate (BER) to analyze the performance of channel estimation and wireless PHY, respectively.

In Fig.~\ref{fig:comparison}(a), we compare the average NMSE of LS, LMMSE, and LSDNN architectures (LSDNN1 and LSDNN2) for a wide range of SNR, and the corresponding BER results on end-to-end wireless transceiver are shown in Fig.~\ref{fig:comparison}(b). We assume Rayleigh fading channel with AWGN. We consider an LSDNN architecture trained on a single SNR referred to as training SNR. Here, the training SNR is selected as 10 dB, and the testing SNR range is from -50 dB to 50 dB. Please refer to Section~\ref{Sec:trainingSNR} for more details on the selection of training SNR.

As shown in Fig.~\ref{fig:comparison}, the performance of the conventional LS and LMMSE architectures matches with their analytical results, and the error for the LMMSE estimator is less than that of the LS estimator. It can be observed that the LMMSE improves the performance of the LS, and the improvement depends on the prior knowledge of the channel and noise statistics. When prior channel statistics are not known accurately, the LMMSE performance degrades, and the NMSE becomes worse than LS at high SNR, as shown in Fig. \ref{fig:comparison}. The BER of the LSDNN1 and LSDNN2  are close to that of the ideal channel estimation approach and significantly outperforms the LS and LMMSE at all SNRs. The NMSE performance of the LSDNN improves with the increase in SNR until the training SNR of 10 dB. Even though LSDNN1 has less complexity, it offers superior performance than LSDNN2. The NMSE is also a function of the number of training samples; hence, the dataset should be sufficiently large. Similar results are also observed for different wireless channels and preamble types. Corresponding plots are not included to avoid the repetition of results.

The improved performance of the LSDNN can be attributed to the capability of the DNN to extract channel features and learn a meaningful representation of the given dataset. Specifically, it accurately learns to distinguish between the AWGN noise and the channel gain. The LS estimation does not consider noise, resulting in noisy estimates at low SNRs. As SNR increases, the noise content in the LS estimation reduces, resulting in improved performance. This is evident from Fig.~\ref{fig:mag} where we have analyzed the plots of estimated channel gain magnitudes at different OFDM sub-carriers for three different SNRs. It can be observed that LSDNN and LMMSE performance is close to ideal channel estimation even at low SNR, with the former better than the latter at all sub-carriers. The erroneous knowledge of channel parameters leads to degradation in LMMSE performance, as shown in  Fig.~\ref{fig:mag}. Such prior knowledge is not needed in the LSDNN.

\section{Channel Estimation on SoC}\label{sec:hwImpl}

In this section, we discuss the algorithm to architecture mapping of three channel estimation approaches, LS, LSDNN, and LMMSE, on FPGA (hardware) part of the ZMPSoC. The software implementation of these algorithms on an ARM processor is straightforward, and corresponding results are given in Section~\ref{sec:results}. We begin with the LSDNN architecture, which includes LS estimation as well.

\begin{figure*}
\centering
\includegraphics[scale=0.625]{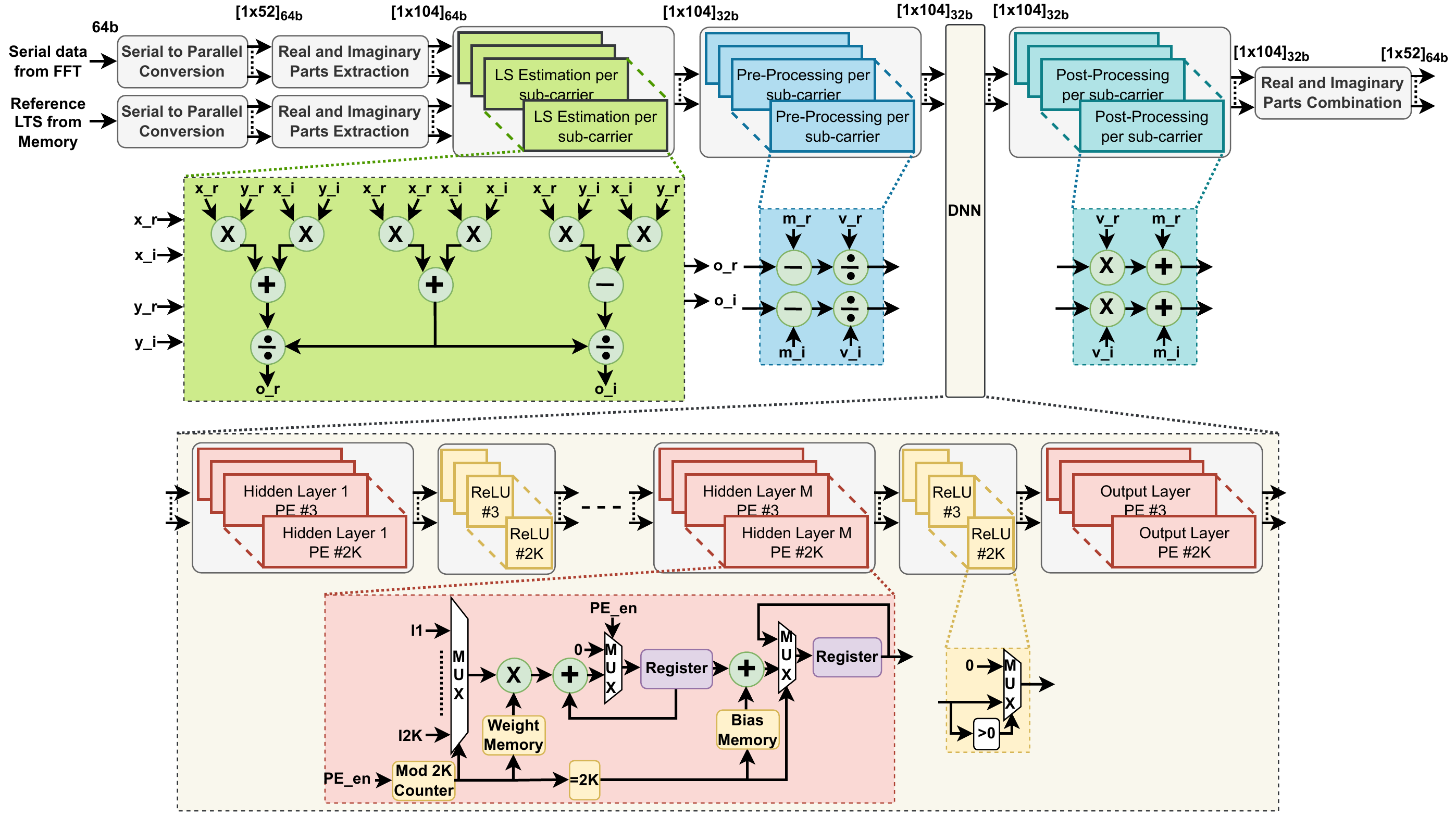}
\caption{\small  Various building blocks of the LSDNN architecture.}
    \label{fig:LSDNN}
\end{figure*}

\subsection{LSDNN Architecture}\label{sub: LSDNN} 

At the receiver, the LTS samples are extracted from the data obtained after the OFDM demodulation, as shown in Fig.~\ref{fig:sysDesign}. LSDNN uses these symbols for channel estimation, and it involves five tasks: 1) Data extraction, 2) LS-based channel estimation, 2) Pre-processing, 3) DNN, and 4) Post-processing, as shown in Fig.~\ref{fig:LSDNN}. In this section, we present the architectures to accomplish these tasks.

The serial data with complex samples received after OFDM demodulation are extracted with independent real and imaginary parts. For the chosen system model with an OFDM symbol comprising of 52 sub-carriers, the input to LS estimation is a vector with 104 real samples. Another input to LS estimation is a reference LTS sequence of the same size. The LS estimation involves the complex division operation of the received ($y_p$) and reference ($x_p$) versions of the LTS. The complex division can be mathematically expressed as shown below, and the corresponding architecture is shown in Fig.~\ref{fig:LSDNN}


\begin{equation}\label{eqn:7}
\begin{aligned}
\hat{\tilde{\ma{H}}}_{\text{LS}} &= \frac{y_p}{x_p} \\
& = \frac{x\_r\times y\_r + x\_i\times y\_i}{x\_r^{2} + x\_i^{2}} + \frac{x\_r\times y\_i + x\_i\times y\_r}{x\_r^{2} + x\_i^{2}} i
 \end{aligned}
\end{equation}
where x\_r, x\_i, y\_r, and y\_i denote the real and imaginary parts of received and reference versions of the LTS, respectively.


The LS estimation of each sub-carrier requires six real multiplications, one real division, three additions, and one subtraction operation. We have explored various architectures of simultaneous calculation of the LS estimation of multiple sub-carriers by appropriate memory partitioning to have parallel access to data and pipelining to improve the utilization efficiency of hardware resources. Please refer to Section~\ref{sec:results} for more details. Further optimizations can be carried out depending on the modulation type of the LTS sequence. \textcolor{black}{For instance, the LS estimation of QPSK modulated LTS sequence needs only one complex multiplication operation. In the case of IEEE 802.11p, the LTS and pilot symbols are BPSK modulated. The LS estimation is reduced to choosing either the received complex value or its two's compliment based on whether the LTS is +1 or -1, respectively.}


The output of the LS channel estimation is further processed using DNN architecture. Before DNN, pre-processing is needed to normalize the DNN input to have zero mean and unit standard deviation. The estimation of mean and standard deviation during inference is not feasible due to limited on-chip memory and latency constraints. Hence, we use the pre-calculated values of the mean and standard deviation using the training dataset. Similarly, the DNN output is de-normalized so that transmitted data can be demodulated for BER-based performance analysis of the end-to-end transceiver. At the architecture level, pre and post-processing incur additional costs in terms of multipliers and adders, as shown in Fig.~\ref{fig:LSDNN}.

\begin{figure}[t]
\centering
\includegraphics[scale=0.48]{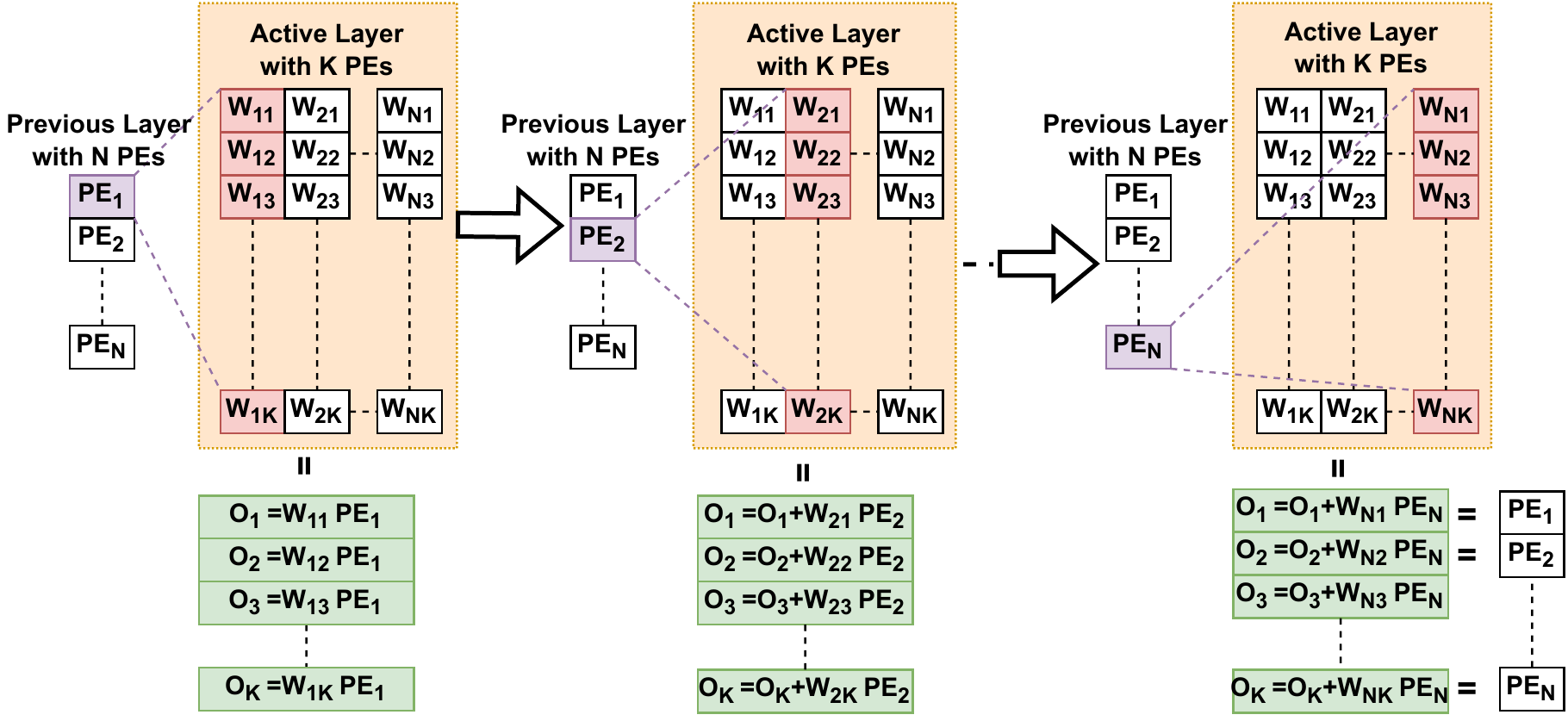}
\caption{\small  Scheduling of operations inside the single DNN layer. }
    \label{fig:sch}
\end{figure}

\begin{figure*}[t]
\centering
\includegraphics[scale=0.7]{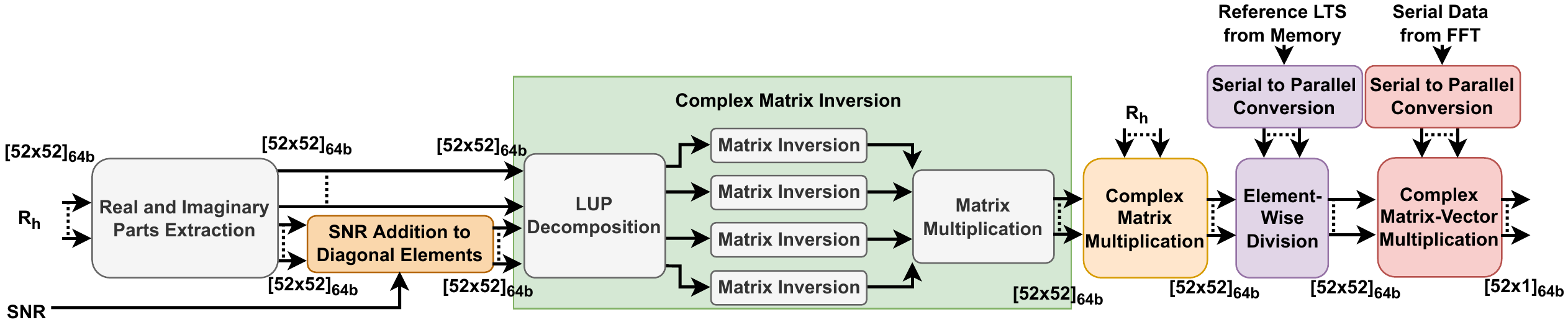}
\caption{\small  Architecture of the LMMSE based channel estimation.}
    \label{fig:mmseEst}
\end{figure*}

As shown in Eq.~\ref{eqn:6} and Fig.~\ref{fig:LSDNN}, each  neuron, i.e. processing element (PE), in DNN consists of a multiplier and adder unit to perform element-wise multiplications between outputs of all PEs of the previous layer and layer-specific weights stored in internal memory. In the end, the output is added with the layer-specific bias term. This means each PE needs to store $N_{l-1}$ weights and one bias value in its memory, where $N_{l-1}$ is the number of PEs in the previous layer. The output of each layer is stored in memory which is completely partitioned to allow parallel access from the PEs in the subsequent layer. The PE functionality is realized sequentially using the counter and multiplexers. Though the parallel realization of all multiplication and addition operations inside the PE is possible due to memory partitioning at PE output, this significantly increases resource utilization and power consumption. ReLU block is added at the end of each hidden layer to introduce non-linearity. It is a hardware-friendly activation function that replaces the negative PE output with zero.
As shown in Fig.~\ref{fig:sch}, the output of the first PE of the previous layer is processed by all PEs in the layer simultaneously, and the output is stored in their respective buffers. Next, the output of the second PE of the previous layer is processed, followed by accumulation with the previous intermediate output in the buffer. This process is repeated until the outputs of all PEs in the previous layer have been processed. After bias addition and ReLU operation, the subsequent layer is activated.

In a completely parallel DNN architecture, all the PEs in a layer are realized using the dedicated hardware resource, and hence the output of all PEs is computed simultaneously. On the other hand, in a completely serial architecture, all PEs in a layer share the hardware resources; hence, the outputs of PEs are computed serially. Depending on resource and latency constraints, various architectures such as fully serial, fully parallel, and serial-parallel architectures for various layers have been explored.  

\begin{figure*}[!t]
\vspace{-0.1cm}
\centering
\includegraphics[scale=0.7]{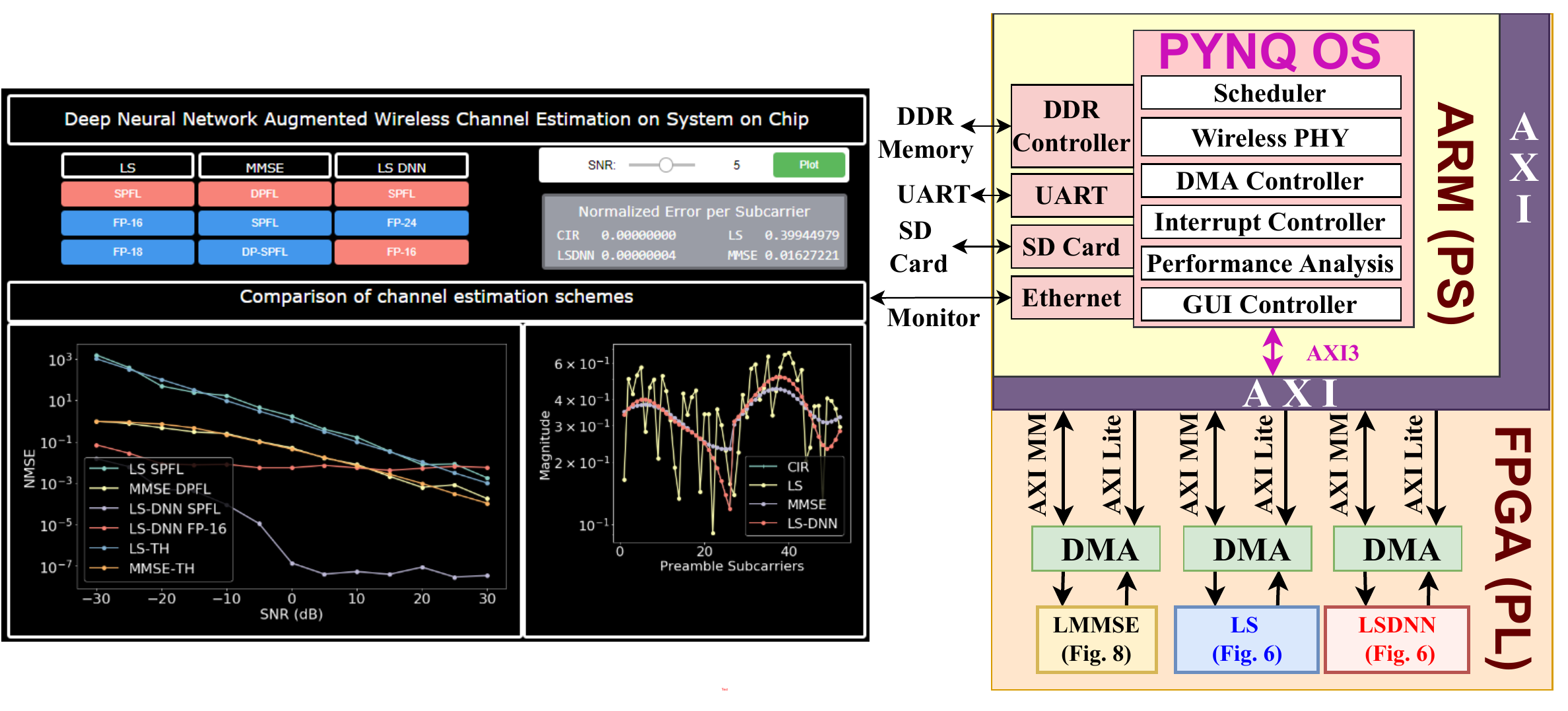}
\vspace{-0.1cm}
\caption{\small Proposed architecture on ZMPSoC and screenshot of Pynq based GUI depicting the real-time demonstration of channel estimation.}
    \label{fig:gui}
\vspace{-0.1cm}
\end{figure*}

\subsection{LMMSE Architecture}\label{subSec:MMSE}
The LMMSE channel estimation requires prior knowledge of the channel correlation matrix ($R_h$) and SNR in addition to LTS symbols. The LMMSE architecture is shown in Fig.~\ref{fig:mmseEst} and is based on Eq.~\ref{eqn:5}. In the beginning, $R_h$ is separated into real and imaginary matrices, and the term (1/SNR) is added with each diagonal element of the real matrix of $R_h$. Then, the inverse of the $R_h$ matrix is performed. This is followed by various matrix multiplication and addition operations. We have modified Xilinx's existing matrix multiplication and matrix inversion reference examples to support the complex number arithmetic since the baseband wireless signal is represented using complex samples. The well-known lower-upper (LU) decomposition method is selected for matrix inversion. We parallelize individual operations like element-wise division and Matrix Multiplication on the FPGA. Every element in the matrix is parallelly processed to compute division, and every row column dot product in matrix multiplication is performed in parallel to speed up the computation. In the end, multiple instances of these IPs are integrated to get the desired LMMSE functionality, as shown in Fig.~\ref{fig:mmseEst}.

\subsection{Hardware Software Co-design and Demonstration}

In Fig.~\ref{fig:gui}, various building blocks of the channel estimation tasks such as LS, LMMSE, LSDNN, scheduler, DMA, interrupt, and GUI controller is shown.  For illustration, we have shown all channel estimation approaches on the FPGA. To enable this, we have developed AXI-stream compatible hardware IPs and interconnected them with PS via direct-memory access (DMA) in the scatter-gather mode for efficient data transfers. Later in Section~\ref{sec:results}, we considered various architectures via hardware-software co-design by moving the blocks between ARM Processor and FPGA. We have deployed PYNQ based operating system on the ARM processor of the {\color{black} ZCU111 platform}, which takes care of various scheduling and controlling operations. It also enables graphical user interface (GUI)  development for real-time demonstration. As shown in  Fig.~\ref{fig:gui}, GUI allows the user to choose various parameters such as channel estimation schemes, SNRs, and word length.

 \begin{figure*}[!b]
        \centering
        \vspace{-0.25cm}
        \includegraphics[scale=0.6]{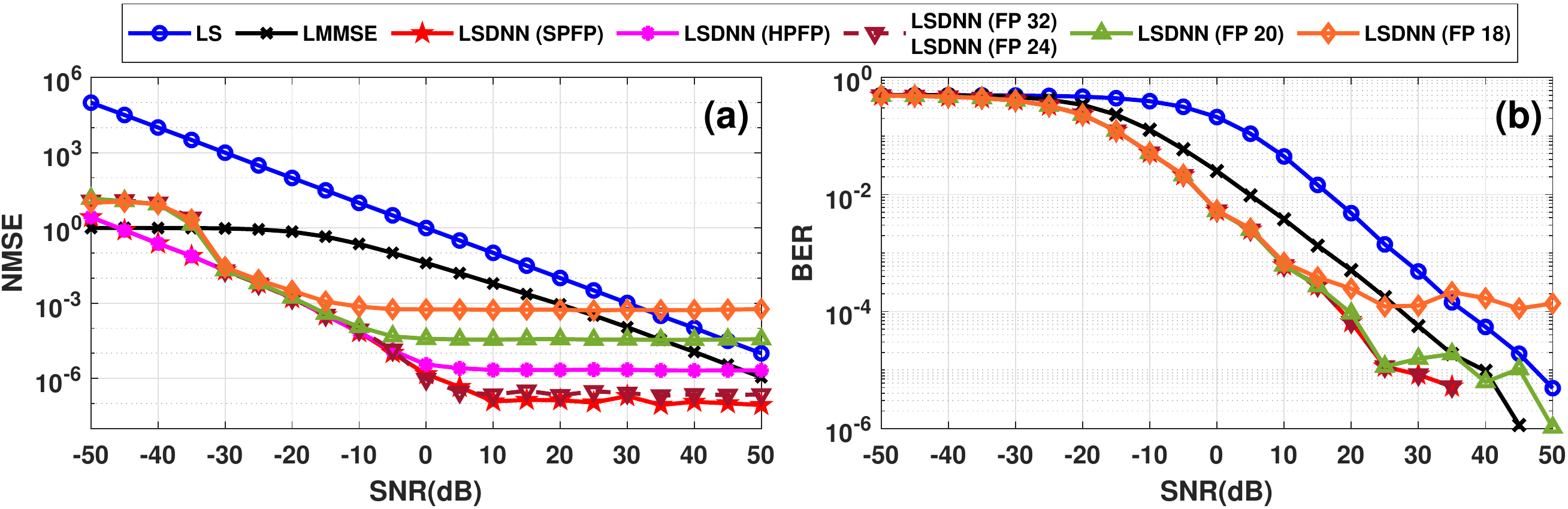}
         \vspace{-0.1cm}
   \caption{(a) NMSE and (b) BER comparisons of various WL architectures of the LSDNN on the ZSoC.}
\label{fig:WL_LSDNN}
        \vspace{-0.2cm}
\end{figure*}

\section{Performance and Complexity Analysis}\label{sec:results}
This section presents the performance analysis of different channel estimation architectures implemented on the {\color{black} Xilinx ZC706 board}
for a wide range of SNRs, word length, and wireless parameters such as channels, preamble type, etc. {\color{black} The FPGA fabric used in the ZC706 board consists of 350 logic cells with 218,600 lookup tables (LUT) and 437,200 flip-flops. It also includes 19.1 Mb of BRAM memory and 900 DSP48 slices.} The resource utilization, execution time, and power consumption of various architectures obtained via word length optimization, hardware-software co-design, and {\color{black} adaptability} are compared. The need for {\color{black} adaptable} architectures and challenges in training DNN for wireless applications are highlighted at the end.


\subsection{Word Length Impact on Channel Estimation Performance}


\begin{table*}[!b]
\centering
\caption{\small Resource utilization and latency comparison of LS, LMMSE, and DNN-Augmented LS channel estimation for different word length implementations.}
\label{tab:table3}
\renewcommand{\arraystretch}{1.2}
 \resizebox{\textwidth}{!}{
\begin{tabular}{|c|c|l|c|c|c|c|c|c|c|}
\hline
\multicolumn{1}{|l|}{\textbf{Sr. No}} & \textbf{Architectures} & \textbf{Word Length} & \multicolumn{1}{l|}{\textbf{Execution time  (ms)}} & \multicolumn{1}{l|}{\textbf{BRAMs}} & \multicolumn{1}{l|}{\textbf{DSPs}} & \multicolumn{1}{l|}{\textbf{LUTs}} & \multicolumn{1}{l|}{\textbf{FFs}} & \multicolumn{1}{l|}{\textbf{PL Power (W)}} & \multicolumn{1}{l|}{\textbf{SoC Power (W)}} \\ \hline
1 & \multirow{3}{*}{\textbf{LS}} & \textbf{SPFL} & 0.0133 & 14 & 96 & 14721 & 15094 & 0.438 & 1.969 \\ \cline{1-1} \cline{3-10} 
2 &  & \textbf{HPFL} & 0.0146 & 5 & 72 & 8371 & 10369 & 0.296 & 1.827 \\ \cline{1-1} \cline{3-10} 
3 &  & \textbf{FP (18,9)} & 0.0155 & 7 & 24 & 14665 & 16649 & 0.373 & 1.904 \\ \hline
4 & \multirow{3}{*}{\textbf{MMSE}} & \textbf{DPFL} & 248.41 & 191.5 & 522 & 44444 & 43485 & 1.158 & 2.689 \\ \cline{1-1} \cline{3-10} 
5 &  & \textbf{SPFL} & 214.7 & 99.5 & 194 & 24249 & 24778 & 0.621 & 2.152 \\ \cline{1-1} \cline{3-10} 
6 &  & \textbf{DP(INV)-SP} & 219.75 & 139.5 & 338 & 36225 & 34898 & 0.918 & 2.449 \\ \hline
7 & \multirow{4}{*}{\textbf{\begin{tabular}[c]{@{}c@{}}LSDNN \\ \\ (104 Parallel PE)\end{tabular}}} & \textbf{SPFL} & 0.0266 & 88 & 322 & 62895 & 58064 & 1.078 & 2.647 \\ \cline{1-1} \cline{3-10} 
8 &  & \textbf{HPFL} & 0.0298 & 32 & 260 & 30436 & 32286 & 0.569 & 2.138 \\ \cline{1-1} \cline{3-10} 
9 &  & \textbf{FP (32,8)} & 0.0186 & 88 & 520 & 157130 & 151544 & 2.46 & 4.029 \\ \cline{1-1} \cline{3-10} 
10 &  & \textbf{FP (24,8)} & 0.0179 & 11 & 260 & 85155 & 82721 & 1.318 & 2.849 \\ \hline
11 & \textbf{\begin{tabular}[c]{@{}c@{}}LSDNN\\ (52 Parellel PE)\end{tabular}} & \textbf{FP (24,8)} & 0.125 & 20 & 32 & 23534 & 23455 & 0.465 & 1.996 \\ \hline
12 & \textbf{\begin{tabular}[c]{@{}c@{}}LSDNN \\ \\ (1 PE)\end{tabular}} & \textbf{FP (24,8)} & 0.236 & 22.5 & 17 & 15450 & 16025 & 0.373 & 1.904 \\ \hline
\end{tabular}%
}
\end{table*}

\begin{table*}[!b]
\centering
\caption{\small Execution times for different Hw/Sw codesign approaches for DNN Augmented LS Estimation}
\label{tab:table4}
\renewcommand{\arraystretch}{1.2}
 \resizebox{\textwidth}{!}{
\begin{tabular}{|l|c|c|c|c|clllcl|}
\hline
\textbf{Sr. No} & \multicolumn{1}{l|}{\textbf{Blocks in PS}} & \multicolumn{1}{l|}{\textbf{Blocks in PL}} & \multicolumn{1}{l|}{\textbf{Execution time (us)}} & \multicolumn{1}{l|}{\textbf{Acceleration factor}} & \multicolumn{1}{l|}{\textbf{BRAM}} & \multicolumn{1}{l|}{\textbf{DSP}} & \multicolumn{1}{l|}{\textbf{LUT}} & \multicolumn{1}{l|}{\textbf{FF}} & \multicolumn{1}{l|}{\textbf{PL Power(W)}} & \textbf{SoC   Power (W)} \\ \hline
1               & \textbf{B1,B2,B3}                          & \textbf{NA}                                & 558                                               & 1                                                 & \multicolumn{6}{c|}{NA}                                                                                                                                                                                             \\ \hline
2               & \textbf{B1,B2}                             & \textbf{B3}                                & 49                                                & 11                                                & \multicolumn{1}{c|}{80}            & \multicolumn{1}{l|}{130}          & \multicolumn{1}{l|}{26622}        & \multicolumn{1}{l|}{25807}       & \multicolumn{1}{c|}{0.453}                & 2.021                   \\ \hline
3               & \textbf{B1}                                & \textbf{B2,B3}                             & 30                                                & 18                                                & \multicolumn{1}{c|}{82}            & \multicolumn{1}{l|}{130}          & \multicolumn{1}{l|}{26456}        & \multicolumn{1}{l|}{28136}       & \multicolumn{1}{c|}{0.453}                & 2.022+                  \\ \hline
4               & \textbf{NA}                                & \textbf{B1,B2,B3}                          & 26                                                & 21                                                & \multicolumn{1}{c|}{88}            & \multicolumn{1}{l|}{322}          & \multicolumn{1}{l|}{62895}        & \multicolumn{1}{l|}{58064}       & \multicolumn{1}{c|}{1.078}                & 2.647                  \\ \hline
\end{tabular}
}
\end{table*}


In Section~\ref{Sec:SimRes}, we have analyzed the performance of various channel estimation approaches on floating-point arithmetic in Matlab. 

{\color{black} Conventionally, floating-point architectures using double-precision floating point (DPFP) and single-precision floating point (SPFP) WLs offer excellent functional accuracy. However, they require high resources, power, and execution time. Since the extremely-large dynamic range offered by floating-point arithmetic may not be needed due to the presence of fixed-width analog-to-digital converters (ADCs), fixed-point WL architectures are preferred. The FPGA offers complete flexibility in the WL selection; detailed analysis is needed to choose the appropriate WL. Here, we use $W$ bits to represent each variable in fixed point WL. Out of $W$ bits, we use  $I$ and ($W-I$) bits to represent the integer and fractional parts, respectively.  To identify appropriate values of $W$ and $I$ for the given architecture, we first use sufficient large $W$ and analyze the functional accuracy for different values of $I$ for all datasets and choose the smallest $I$, that offers the same performance as that of SPFL architecture. For the selected $I$, we find the lowest $W$ using the same process.}

In Fig.~\ref{fig:WL_LSDNN}(a) and (b), we compare the NMSE and BER performance of the various LSDNN architectures, respectively, for six different WLs along with the SPFP architecture of the LS and DPFP architecture of the LMMSE. The architectures with the half-precision floating point (HPFL) and fixed-point architectures with WL below 24 {\color{black}and integer length of 8 (FP(24,8))} lead to significant degradation in the NMSE. However, degradation in NMSE may not correspond to degradation in the BER due to data modulation in the wireless transceiver. This is evident from Fig.~\ref{fig:WL_LSDNN}(b), where the architecture with HPFL offers the BER performance same as that of the SPFL architecture. On the other hand, the BER of the fixed-point architecture with WL below {\color{black} FP(24,8)} leads to significant degradation in the performance. 


 \begin{figure}[!b]
        \centering
        \vspace{-0.25cm}
        \includegraphics[scale=0.65]{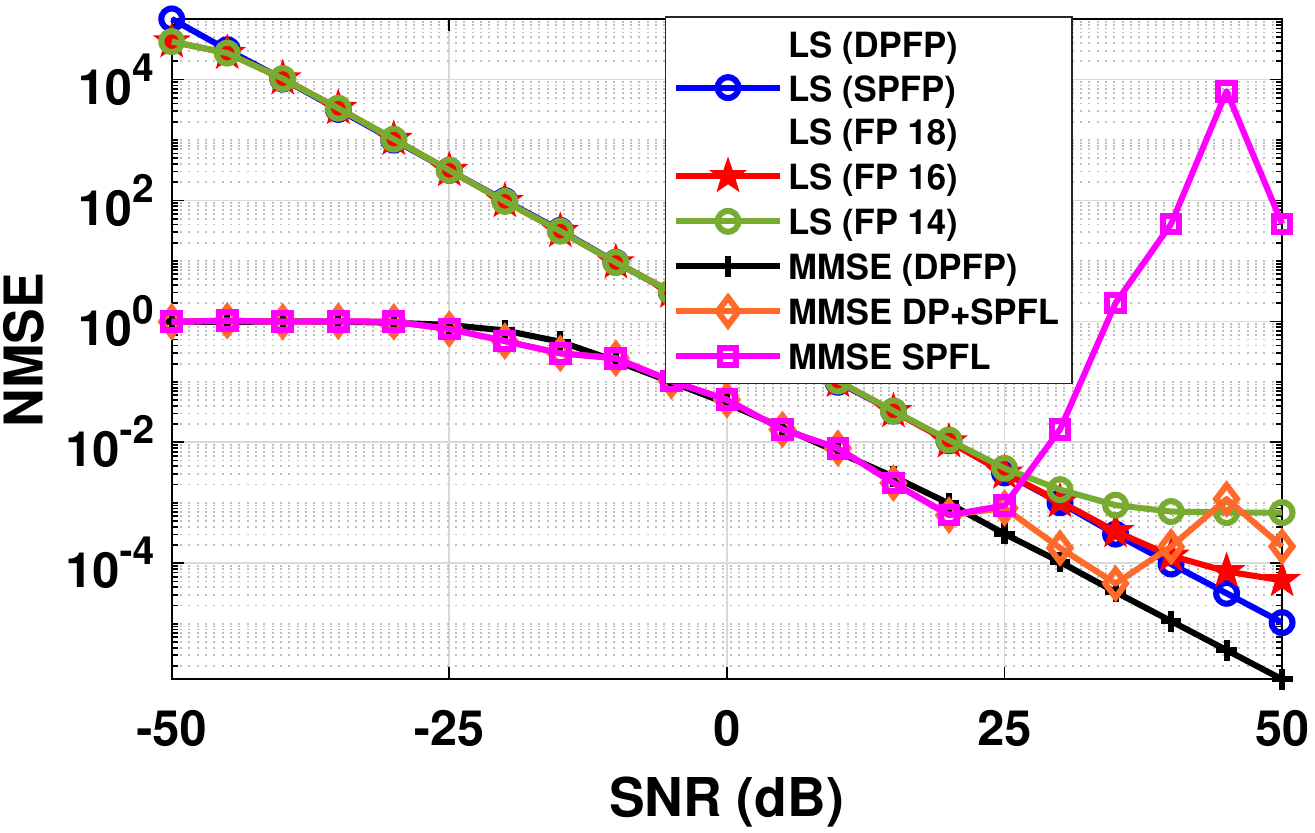}
         \vspace{-0.1cm}
   \caption{{\color{black}NMSE Comparison of various WL architectures of (a) LS and (b) LMMSE on the ZSoC.}}
\label{fig:WL_LS}
        \vspace{-0.2cm}
\end{figure}
\setcounter{figure}{12}
 \begin{figure*}[!b]
        \centering
        \vspace{-0.25cm}
       \includegraphics[scale=0.6]{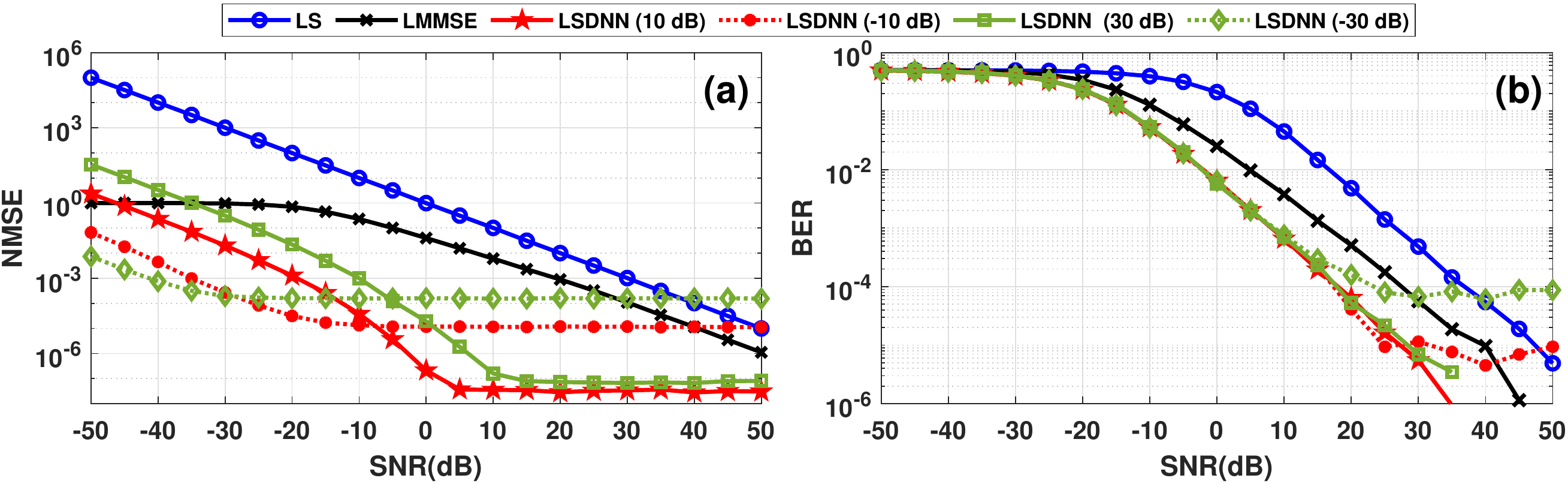}
       
         \vspace{-0.1cm}
   \caption{(a) NMSE and (b) BER comparisons of various LSDNN architectures trained on different training SNRs.}
\label{fig:trainVariation}
        \vspace{-0.2cm}
\end{figure*}

The WL selection is architecture-dependent; hence, the WL for the LS may not be the same as that of LSDNN. Based on the detailed experimental analysis, the LS architecture with fixed-point WL of 18 bits {\color{blue} (FP(18,9))} or higher offers the same performance as its SPFL architecture, as shown in Fig~\ref{fig:WL_LS}(a). In the case of LMMSE, fixed-point architecture is not feasible. Even SPFL architecture fails to offer the desired performance, especially at high SNRs; hence, DPFL is needed. Further analysis revealed that only the matrix inverse sub-block needs DPFL WL while the rest of the sub-blocks can be realized in SPFL WL as shown in Fig.~\ref{fig:WL_LS}(b). Such analysis highlights the importance of experiments on the SoC for different WL since such results can not be obtained using simulations.

\begin{table*}[!t]
\centering
\caption{\small Resource utilization and latency comparison of LSDNN for wireless PHY with different FFT size}
\label{tab:table5}
\renewcommand{\arraystretch}{1.2}
 \resizebox{\textwidth}{!}{
 \begin{tabular}{|l|l|l|l|l|l|l|l|l|l|}
\hline
\textbf{Sr. No} & \textbf{FFT Size} & \textbf{Execution time (us)} & \textbf{Acceleration factor} & \textbf{BRAM} & \textbf{DSP} & \textbf{LUT} & \textbf{FF} & \textbf{PL Power(W)} & \textbf{SoC   Power (W)} \\ \hline
1               & \textbf{64}       & 26                           & 21                           & 88            & 322          & 62895        & 58064       & 1.078                & 2.647                  \\ \hline
2               & \textbf{128}      & 48                           & 28                           & 195           & 512          & 83193        & 80679       & 1.474                & 3.005                   \\ \hline
3               & \textbf{256}      & 85                           & 31                           & 339           & 832          & 119043       & 118606      & 2.206                & 3.737                   \\ \hline
\end{tabular}
}
\end{table*}

\subsection{Resource, Power and Execution Time Comparison}
In this section, we compare the resource utilization, execution time, and power consumption of various architectures on the ZSoC platform. 
In Table~\ref{tab:table3}, we consider three different WL architectures of the LS and LMMSE and six different WL architectures of the LSDNN. It is assumed that the complete architecture is realized in the FPGA (PL) part of the SoC. The execution time of the LS architecture is the lowest, while the execution time of the LMMSE architecture is the highest. The execution time of the LSDNN is significantly lower than that of LMMSE and is around 1.5-2 times that of the LS architecture. Similarly, resource utilization and power consumption of the LSDNN are higher than that of the LS. This is the penalty paid to gain significant improvement in channel estimation performance. However, the LSDNN offers a significantly lower execution time than the LMMSE, and resource utilization is also lower, especially in terms of BRAM and DSP units which are limited in numbers on FPGA. The use of fixed-point architectures leads to significant improvement in all three parameters. In case of LSDNN, we also considered two additional architectures: 1) Serial-parallel architecture where 2 PEs share the same hardware resources, and 2) Fully serial architecture where all PEs share the same hardware resources. As we move from a completely parallel architecture to these two architectures, we can gain significant savings in the number of resources, especially DSPs. However, the execution time increases, which is still significantly lower than the LMMSE. Further improvement is possible if the input data type of the architecture is changed to fixed-point so that fixed-to-floating point conversion overhead at the input and output can be removed.

{\color{black}Next, we consider various LSDNN architectures obtained via hardware-software co-design. For this purpose, LSDNN is divided into three blocks – LS (B1), normalization and denormalization (B2), and DNN (B3). In Table~\ref{tab:table4}, we consider four architectures: 1) B1, B2, B3 in PS, 2) B1, B2 in PS, and B3 in PL, 3) B1 in PS, and B2, B3 in PL, and 4) B1, B2, B3 in PL (Same as Table~\ref{tab:table3}). }
{\color{black} Moving DNN to PL gives $11\times$ improvement in execution time with respect to PS implementation as FPGA can exploit inherent parallelism in DNNs, as explained in section \ref{sub: LSDNN}, to speed up the processing. Normalization and denormalization can also be performed in parallel as there is no data dependency for these operations and thus can be accelerated by moving to PL. Execution time can further be reduced by moving LS to PL and processing multiple subcarriers in parallel, thus achieving an overall acceleration factor of $21\times$ when all the data processing blocks are implemented in PL compared to complete PS implementation.
Reducing execution time by moving blocks to PL and introducing parallelism increases the FPGA fabric's resource utilization and power consumption. Thus, when choosing a particular architecture, there is a trade-off between latency and resource utilization/power consumption. }

With the evolution of wireless networks, the transmission bandwidth and hence, the FFT size of the OFDM PHY is also increasing. The increase in FFT size leads to an increase in the complexity of the channel estimation architecture, which in turn demands efficient acceleration via FPGA on the SoC. In Table~\ref{tab:table5}, we compare the effect on execution time and corresponding acceleration factor for different FFT sizes. It can be observed that the gain in acceleration factor increases with the increase in the FFT size mainly due to more opportunities for parallel arithmetic operations. Thus, LSDNN architecture has the potential to offer a significant improvement in performance and execution time for next-generation wireless PHY.

\subsection{Effect of Training SNR}
\label{Sec:trainingSNR}

For the results presented till now, the LSDNN model is obtained using the training SNR of 10 dB. Training a DNN on a single SNR allows it to converge faster and model the channel better as the DNN does not have to encounter multiple noise distributions during the training phase, thus reducing training complexity. However, faster training should not compromise the functional performance of the LSDNN. For in-depth performance analysis, we analyze the effect of training SNR on the performance of the LSDNN model.

\setcounter{figure}{11}
\begin{figure}[!b]
\centering
\includegraphics[scale=0.6]{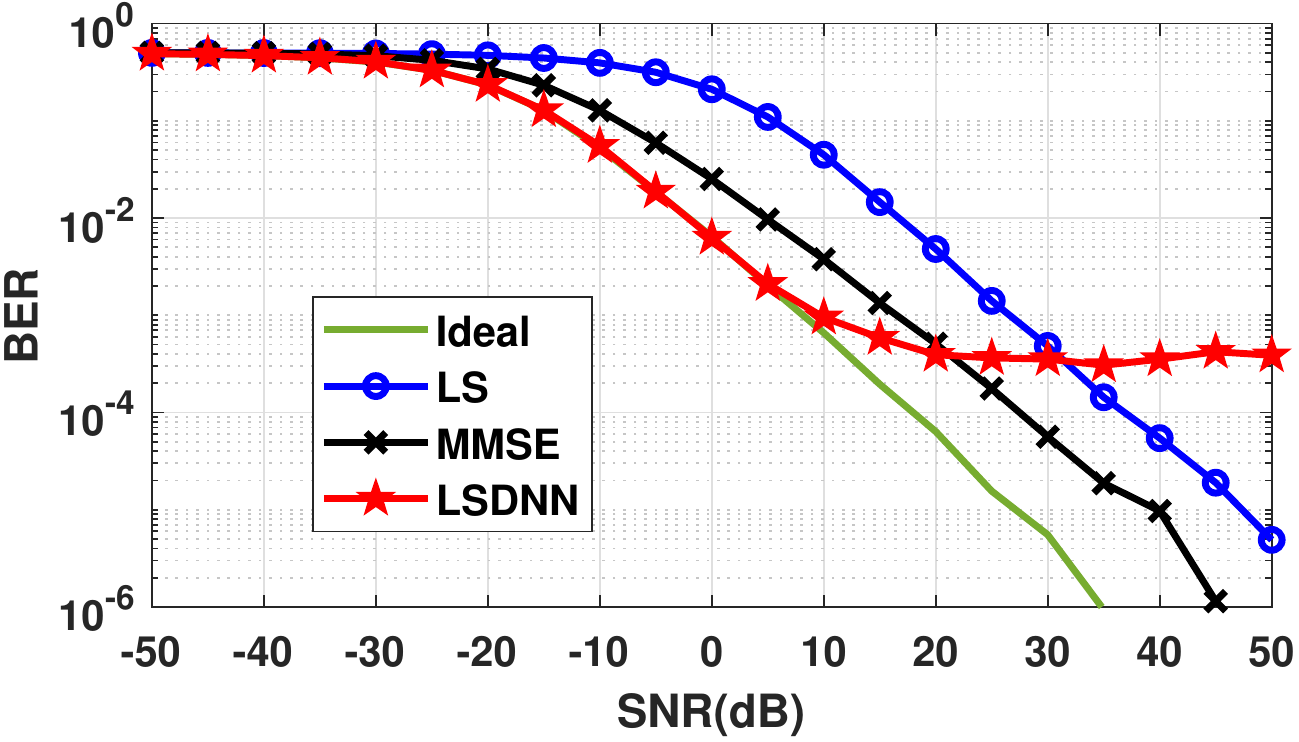}
\caption{BER performance of the LSDNN architecture trained with dataset containing samples from complete SNR range of -50dB to 50dB.}
\label{fig:allSNR}
\end{figure}
\setcounter{figure}{14}
 \begin{figure*}[!b]
        \centering
        \vspace{-0.25cm}
       \includegraphics[scale=0.6]{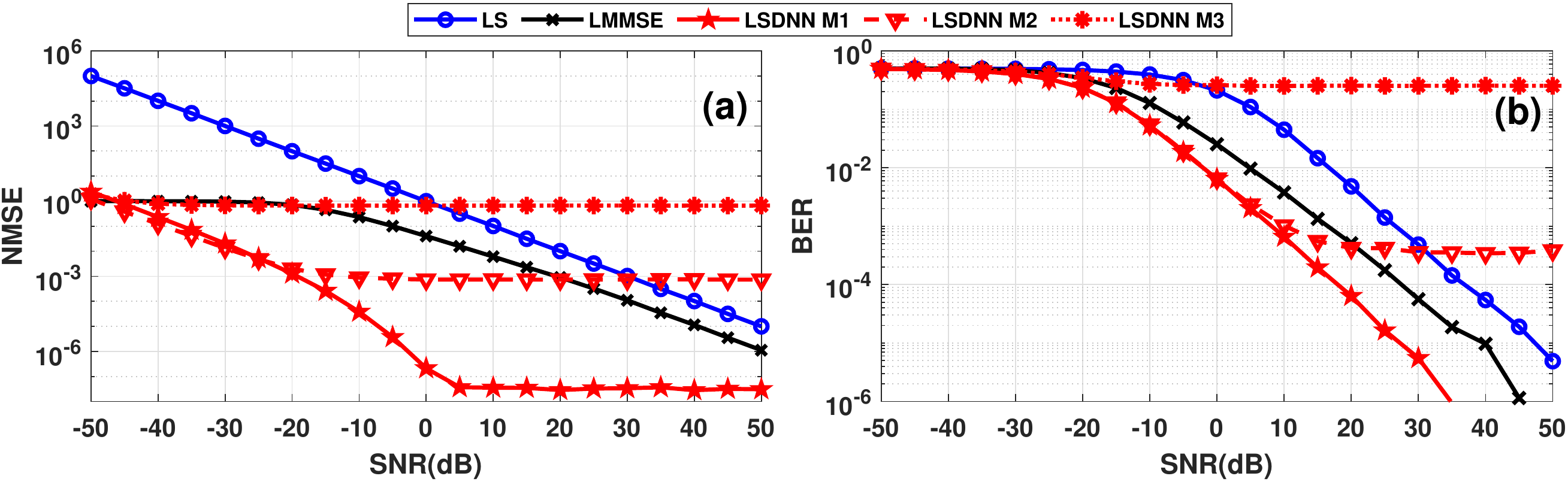}
         \vspace{-0.1cm}
   \caption{(a) NMSE and (b) BER comparisons of various channel estimation approaches on different models. The LSDNN model is trained on model M1.}
\label{fig:chMod}
        \vspace{-0.2cm}
\end{figure*}

We initially consider various architectures of the LSDNN model trained using the dataset comprising all SNR samples ranging from -50 dB to 50 dB. As shown in Fig.\ref{fig:allSNR}, the NMSE and BER performance is poor due to the reasons mentioned before. However, the selection of training SNR is not trivial since optimal training SNR may vary depending on the testing SNR, i.e., the deployment environment. For instance, as shown in Fig.~\ref{fig:trainVariation} (a) and (b), single training SNR affects the performance of LSDNN considerably, especially at high testing SNRs. To address this issue, we consider the selection of training SNR based on the testing SNR range, and corresponding results are shown in Fig.~\ref{fig:trainSnr}. It shows the NMSE for different testing SNRs plotted against the selected training SNR. For each testing SNR, the NMSE first decreases as the training dataset SNR is increased till it reaches its minimum value and then starts increasing. The best training SNR is where the NMSE is minimum, e.g., for the testing SNR range of 0 dB - 20dB, the lowest NMSE corresponds to a training SNR of 10dB. Similarly, for testing SNR of -10 dB to 10 dB, the optimal training SNR is 0 dB. Note that training SNR is irrelevant at low SNRs due to the significant impact of noise on the received signal, and DNN is unable to learn the channel properties, resulting in high NMSE and BER. Such dependence on the testing SNR demands reconfigurable architecture that can switch between various LSDNN models depending on the given testing SNR range. Such architecture is presented in the next section.

\setcounter{figure}{13}
\begin{figure}[!t]
\centering
\includegraphics[scale=0.6]{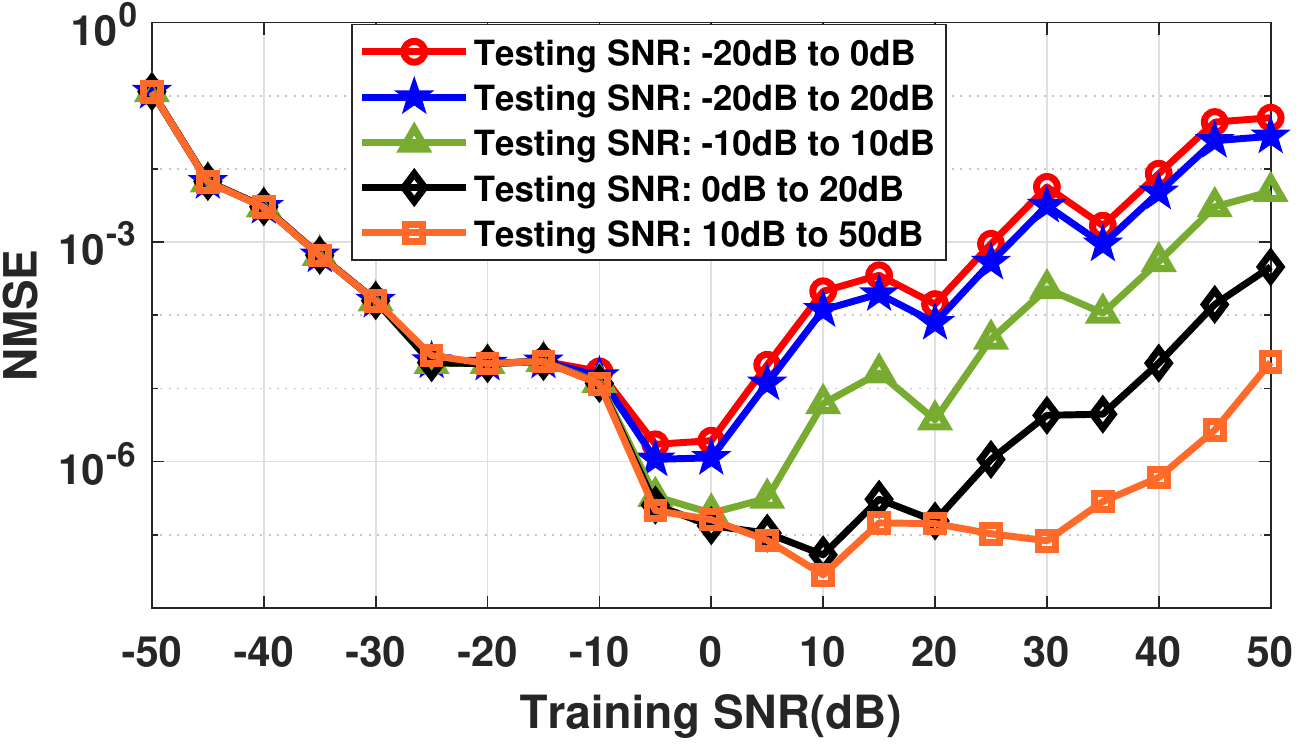}
\caption{\small Comparison of effect of training SNR on the NMSE for various ranges of the testing SNR.}
    \label{fig:trainSnr}
\end{figure}




\subsection{{\color{black} Adaptable} Architecture}
\label{Sec:Reconfarch}
Though the LSDNN approach offers better performance than LMMSE without the need of prior channel knowledge, LSDNN performance depends on the efficacy of the training dataset with respect to the validation environment. For instance, the performance of LSDNN may degrade if the parameters such as testing SNR, preamble type, pilot type, etc., of the training and validation environment do not match. For instance, the performance of the LSDNN trained on model $M1$ degrades significantly on two different models, $M2$ and $M3$, as shown in Fig.~\ref{fig:chMod}. This is expected because for a DNN to perform satisfactorily, the training and testing conditions should remain the same. To address this challenge, we need an {\color{black} adaptable} architecture for LSDNN to enable an on-the-fly switch between various LSDNN models.

We propose an {\color{black} adaptable} architecture for LSDNN where the parameters of various LSDNN-trained models are pre-computed and stored in memory. Here, it is assumed that the DNN architecture is fixed for various channel conditions. Thus, the same hardware architecture can be used across multiple testing environments. {\color{black} In Table~\ref{tab:table6},we consider two architectures based on the type of memory in the SoC. In the first architecture, external DDR memory stores the model parameters, and the parameters of the selected LSDNN model are transferred to the PL via DMA. Such architecture suffers from high execution time since the parameters are read from external DDR memory. However, on-field {\color{black} adaptability} is possible since the parameters corresponding to new channel conditions can be added in DDR using PS without needing hardware reconfiguration. The low cost of the DDR allows the storage of a large number of DNN parameters. In the second approach, model parameters are stored in the BRAM of the PL, resulting in lower execution time. Since the BRAM is dual-port memory, we can perform some of the operations in parallel, thereby significantly reducing the execution time at the cost of a slight increase in resource utilization. Due to limited BRAM, the second approach limits the number of supported DNN models and hence, different channel conditions. The support for new channel conditions demands updation of BRAM content which requires FPGA configuration either completely or partially. Complete reconfiguration is needed when DNN architecture is different and partial reconfiguration of BRAM is needed when DNN architecture is the same, but parameters are different. Note that FPGA reconfiguration can still be done remotely without requiring product recall. As part of future works, the second architecture can be enhanced to optimize BRAM utilization via dynamic partial reconfiguration, thereby making the BRAM utilization independent of the number of models. Such reconfigurable architecture also needs intelligence to detect the change in the environment and reconfigure the hardware. The design of intelligent and reconfigurable channel estimation architecture is an important research direction.}

\begin{table}[!t]

\centering
\caption{\small Comparison of various {\color{blue} adaptable} LSDNN architectures}
\label{tab:table6}

\begin{tabular}{|l|l|ll|}
\hline
\multicolumn{1}{|c|}{\multirow{2}{*}{\textbf{}}} & \multicolumn{1}{c|}{\textbf{External DDR }} & \multicolumn{2}{c|}{\textbf{On-chip BRAM}}                \\ \cline{3-4} 
\multicolumn{1}{|c|}{}                           & \multicolumn{1}{c|}{\textbf{Memory}}                                      & \multicolumn{1}{l|}{\textbf{4 models}} & \textbf{8 models} \\ \hline
\textbf{Execution time   (ms)}                   & 0.257                                                      & \multicolumn{1}{l|}{0.0264}            & 0.029             \\ \hline
\textbf{LUT}                                     & 65926                                                      & \multicolumn{1}{l|}{68424}             & 70634             \\ \hline
\textbf{FF}                                      & 70300                                                      & \multicolumn{1}{l|}{63802}             & 63814             \\ \hline
\textbf{BRAM}                                    & 92                                                         & \multicolumn{1}{l|}{221.5}             & 253.5             \\ \hline
\textbf{DSP}                                     & 300                                                        & \multicolumn{1}{l|}{324}               & 324               \\ \hline
\textbf{PL power (W)}                                & 0.854                                                      & \multicolumn{1}{l|}{1.099}             & 1.121             \\ \hline
\textbf{SoC power (W)}                               & 2.423                                                      & \multicolumn{1}{l|}{2.668}             &   2.69           \\ \hline
\end{tabular}
\end{table}

\section{ASIC Implementation}\label{sec:ASIC Implementation}
In addition to FPGA, ASIC is another popular hardware platform for wireless PHY. In this section, we assess the PPA delivered by various architectures in the ASIC implementation using  45 nm CMOS technology.
First, we synthesize the Verilog code used for FPGA implementation, using suitable constraints, to obtain the gate level netlist. We verify the functional correctness of the design using a combinational equivalence checker and ensure its temporal safety using static timing analysis. The validated gate-level netlist is converted into the final Graphic Data System (GDS) using the physical design steps such as floor-planning, placement, clock-tree synthesis, and routing. Finally, we carry out various sign-off checks on the ASIC implementation using physical verification and timing analysis tools. We report the PPA of the implementations with various channel estimation architectures in Fig.~\ref{fig:topArch}. To obtain the maximum operable frequency for a given architecture, we have carried out an ASIC implementation flow with multiple timing constraints. We gradually increased the target frequency until no more improvement was possible.

We found that for 32-bit implementation, LS achieves the operating frequency of 350 MHz, while the LSDNN achieves the operating frequency of 200 MHz, and the LSDNN implementation consumes 4.8$\times$ more area and 2.5$\times$ higher power compared to the LS implementation at their corresponding frequencies of operation. These results are expected because LSDNN architecture performs more computation than the corresponding LS implementation.

\setcounter{figure}{15}
\begin{figure}[t]
\centering
\includegraphics[scale = 0.37]{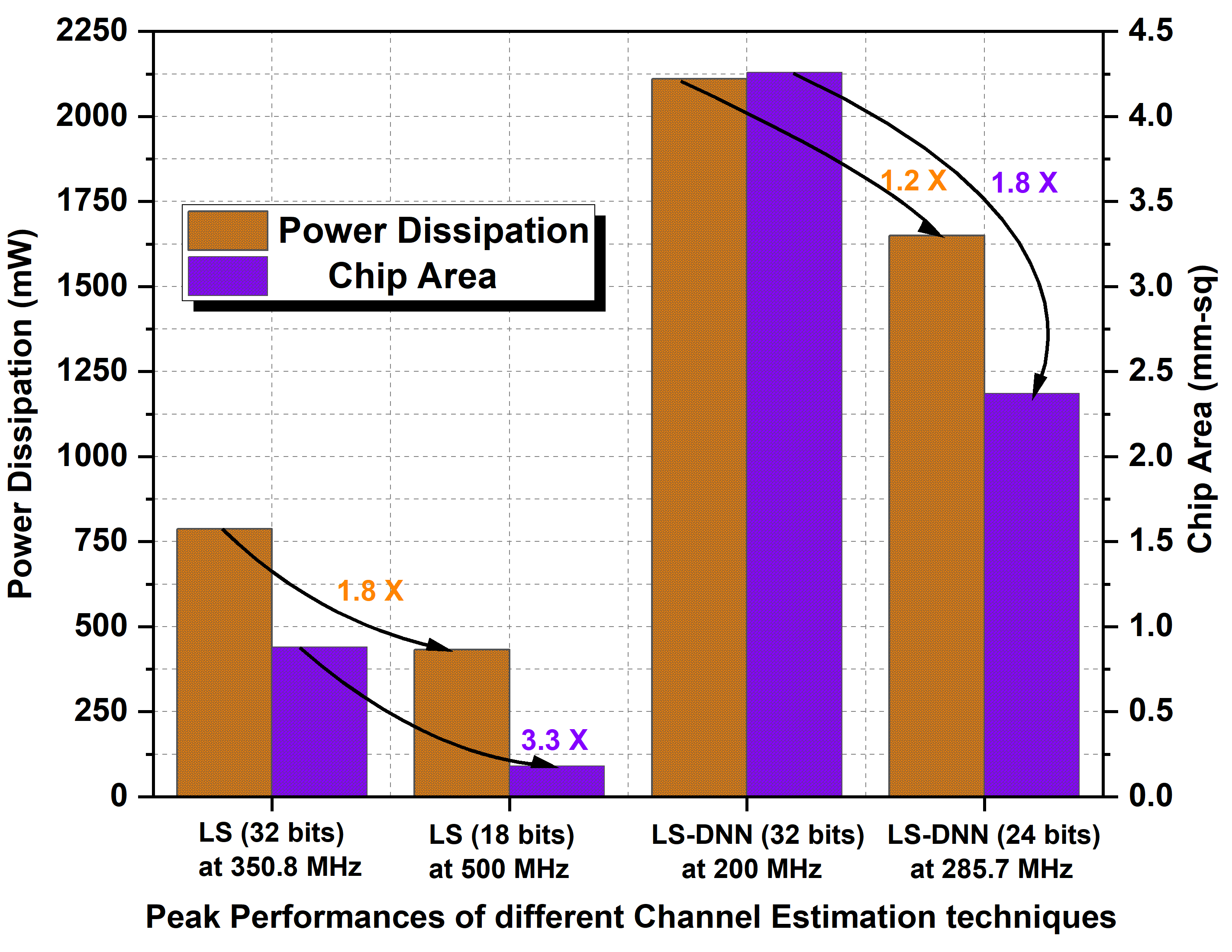}
\caption{\small  Comparative analysis of PPA in ASIC implementations of channel estimation techniques (Vdd = 1V, technology = 45 nm) }
    \label{fig:topArch}
\end{figure}

Further, we study the effect of word-length optimizations on the ASIC implementations of different channel estimation techniques. We found that upon word-length optimization of LS from 32 bits to 18 bits, the 18-bits LS achieves 1.5$\times$ higher operating frequency compared to 32-bits LS (from 350 MHz. to 500 MHz.). Moreover, the 18-bits LS takes 3.3$\times$ less area and consumes 1.8$\times$ less power than the 32-bits LS at their peak operating frequencies.
Likewise, upon word-length optimization of LSDNN from 32 bits to 24 bits, the 24-bits LSDNN achieves 1.4$\times$ higher operating frequency compared to 32-bits LSDNN (from 200 MHz to 285.7 MHz). Moreover, the 24-bits LSDNN takes 1.8$\times$ less area and 1.2$\times$ less power compared to the 32-bits LSDNN at their peak operating frequencies. Thus, the word-length optimization delivers PPA improvements in ASIC implementations also, both for the LS and LSDNN architectures.



\section{Conclusions and Future Directions}\label{sec:conclusions}
In this work, we proposed a deep neural network (DNN) augmented least-square (LSDNN) based channel estimation for wireless physical layer (PHY). We have compared the performance, resource utilization, power consumption, and execution time of LSDNN with conventional LS and linear minimum mean square estimation (LMMSE) approach on Zynq multiprocessor system-on-chip (ZMPSoC) and application-specific integrated circuits (ASIC) platforms. Numerous experiments validate the superiority of the LSDNN over LS and LMMSE. The AXI-compatible hardware IPs and PYNQ-based graphical user interface (GUI) demonstrating the real-time performance comparison on the ZMPSoC are a physical deliverable of this work.

Future works include an extension of the proposed approach for cellular OFDM PHY with comb-based pilots instead of the preamble in IEEE 802.11. The proposed approach can offer a significant reduction in complexity for millimeter-wave massive MIMO networks. The demonstration of the proposed approach in a real radio environment is critical for commercial deployment. 



\section*{Acknowledgement}
We thank Mr. Mohammed Sajjad Jafri (Research Intern, IIIT Delhi) for his assistance with developing GUI to demonstrate the proposed work on ZMPSoC.

\bibliography{References.bib}

\begin{thebibliography}{10}

\bibitem{Zeadally2020ITSstandard}
S.~Zeadally, M.~A. Javed, and E.~B. Hamida, ``Vehicular communications for its:
  Standardization and challenges,'' {\em IEEE Communications Standards
  Magazine}, vol.~4, no.~1, pp.~11--17, 2020.

\bibitem{Mumtaz2021lowLatency6g}
S.~Mumtaz, V.~G. Menon, and M.~I. Ashraf, ``Guest editorial: Ultra-low-latency
  and reliable communications for 6g networks,'' {\em IEEE Communications
  Standards Magazine}, vol.~5, no.~2, pp.~10--11, 2021.

\bibitem{6814271}
Y.~Liu, Z.~Tan, H.~Hu, L.~J. Cimini, and G.~Y. Li, ``Channel estimation for
  ofdm,'' {\em IEEE Communications Surveys \& Tutorials}, vol.~16, no.~4,
  pp.~1891--1908, 2014.

\bibitem{7447802}
K.~K. Nagalapur, F.~Brännström, E.~G. Ström, F.~Undi, and K.~Mahler, ``An
  802.11p cross-layered pilot scheme for time- and frequency-varying channels
  and its hardware implementation,'' {\em IEEE Transactions on Vehicular
  Technology}, vol.~65, no.~6, pp.~3917--3928, 2016.

\bibitem{van1995channel}
Z.~Yuanjin, ``A novel channel estimation and tracking method for wireless ofdm
  systems based on pilots and kalman filtering,'' in {\em 2003 IEEE
  International Symposium on Circuits and Systems (ISCAS)}, vol.~2, pp.~II--II,
  2003.

\bibitem{9247524}
F.~Restuccia and T.~Melodia, ``Deep learning at the physical layer: System
  challenges and applications to 5g and beyond,'' {\em IEEE Communications
  Magazine}, vol.~58, no.~10, pp.~58--64, 2020.

\bibitem{DLPHY2}
J.~Liu, J.~Chen, S.~Luo, S.~Li, and S.~Fu, ``Deep learning driven
  non-orthogonal precoding for millimeter wave communications,'' {\em IEEE
  Journal on Emerging and Selected Topics in Circuits and Systems}, vol.~10,
  no.~2, pp.~164--176, 2020.

\bibitem{yang2019deep}
Y.~Yang, F.~Gao, X.~Ma, and S.~Zhang, ``Deep learning-based channel estimation
  for doubly selective fading channels,'' {\em IEEE Access}, vol.~7,
  pp.~36579--36589, 2019.

\bibitem{wei2020deep}
X.~Wei, C.~Hu, and L.~Dai, ``Deep learning for beamspace channel estimation in
  millimeter-wave massive mimo systems,'' {\em IEEE Transactions on
  Communications}, vol.~69, no.~1, pp.~182--193, 2020.

\bibitem{ma2020data}
X.~Ma and Z.~Gao, ``Data-driven deep learning to design pilot and channel
  estimator for massive mimo,'' {\em IEEE Transactions on Vehicular
  Technology}, vol.~69, no.~5, pp.~5677--5682, 2020.

\bibitem{gizzini2020enhancing}
A.~K. Gizzini, M.~Chafii, A.~Nimr, and G.~Fettweis, ``Enhancing least square
  channel estimation using deep learning,'' in {\em 2020 IEEE 91st Vehicular
  Technology Conference (VTC2020-Spring)}, pp.~1--5, IEEE, 2020.

\bibitem{gizzini2020deep}
A.~K. Gizzini, M.~Chafii, A.~Nimr, and G.~Fettweis, ``Deep learning based
  channel estimation schemes for ieee 802.11 p standard,'' {\em IEEE Access},
  vol.~8, pp.~113751--113765, 2020.

\bibitem{pan2021channel}
J.~Pan, H.~Shan, R.~Li, Y.~Wu, W.~Wu, and T.~Q. Quek, ``Channel estimation
  based on deep learning in vehicle-to-everything environments,'' {\em IEEE
  Communications Letters}, vol.~25, no.~6, pp.~1891--1895, 2021.

\bibitem{DLPHY1}
C.-F. Teng and A.-Y. Wu, ``A 7.8–13.6 pj/b ultra-low latency and
  reconfigurable neural network-assisted polar decoder with multi-code length
  support,'' {\em IEEE Transactions on Circuits and Systems I: Regular Papers},
  vol.~68, no.~5, pp.~1956--1965, 2021.

\bibitem{DL_Rohith}
R.~Rajesh, S.~J. Darak, A.~Jain, S.~Chandhok, and A.~Sharma,
  ``Hardware–software co-design of statistical and deep-learning frameworks
  for wideband sensing on zynq system on chip,'' {\em IEEE Transactions on Very
  Large Scale Integration (VLSI) Systems}, vol.~31, no.~1, pp.~79--89, 2023.

\bibitem{8839651}
D.~Gündüz, P.~de~Kerret, N.~D. Sidiropoulos, D.~Gesbert, C.~R. Murthy, and
  M.~van~der Schaar, ``Machine learning in the air,'' {\em IEEE Journal on
  Selected Areas in Communications}, vol.~37, no.~10, pp.~2184--2199, 2019.

\bibitem{8233654}
T.~Wang, C.-K. Wen, H.~Wang, F.~Gao, T.~Jiang, and S.~Jin, ``Deep learning for
  wireless physical layer: Opportunities and challenges,'' {\em China
  Communications}, vol.~14, no.~11, pp.~92--111, 2017.

\bibitem{9508930}
S.~Zhang, J.~Liu, T.~K. Rodrigues, and N.~Kato, ``Deep learning techniques for
  advancing 6g communications in the physical layer,'' {\em IEEE Wireless
  Communications}, vol.~28, no.~5, pp.~141--147, 2021.

\bibitem{9440456}
K.~Mei, J.~Liu, X.~Zhang, N.~Rajatheva, and J.~Wei, ``Performance analysis on
  machine learning-based channel estimation,'' {\em IEEE Transactions on
  Communications}, vol.~69, no.~8, pp.~5183--5193, 2021.

\bibitem{8640815}
M.~Soltani, V.~Pourahmadi, A.~Mirzaei, and H.~Sheikhzadeh, ``Deep
  learning-based channel estimation,'' {\em IEEE Communications Letters},
  vol.~23, no.~4, pp.~652--655, 2019.

\bibitem{8944280}
L.~Li, H.~Chen, H.-H. Chang, and L.~Liu, ``Deep residual learning meets ofdm
  channel estimation,'' {\em IEEE Wireless Communications Letters}, vol.~9,
  no.~5, pp.~615--618, 2020.

\bibitem{code}
S.~A.~U. Haq, ``{Demonstration video. Source codes and datasets will be shared
  after paper acceptance.}.''
  \href{https://drive.google.com/drive/folders/1qE3Mq2fOF78t_AQlv9tbBnniIKYCky-3?usp=share_link}{Link},
  2022.
\newblock [Online; accessed 5-January-2023].

\bibitem{shannon1948mathematical}
C.~E. Shannon, ``A mathematical theory of communication,'' {\em The Bell system
  technical journal}, vol.~27, no.~3, pp.~379--423, 1948.

\bibitem{kay1998fundamentals}
S.~M. Kay, ``Fundamentals of statistical signal processing. detection theory,
  volume ii,'' {\em Printice Hall PTR}, vol.~1545, 1998.

\bibitem{karray2013queueing}
M.~K. Karray and M.~Jovanovic, ``A queueing theoretic approach to the
  dimensioning of wireless cellular networks serving variable-bit-rate calls,''
  {\em IEEE Transactions on Vehicular Technology}, vol.~62, no.~6,
  pp.~2713--2723, 2013.

\bibitem{singh2021toward}
N.~Singh, S.~S. Santosh, and S.~J. Darak, ``Toward intelligent reconfigurable
  wireless physical layer (phy),'' {\em IEEE Open Journal of Circuits and
  Systems}, vol.~2, pp.~226--240, 2021.

\bibitem{darak2019multi}
S.~J. Darak and M.~K. Hanawal, ``Multi-player multi-armed bandits for stable
  allocation in heterogeneous ad-hoc networks,'' {\em IEEE Journal on Selected
  Areas in Communications}, vol.~37, no.~10, pp.~2350--2363, 2019.

\bibitem{CESoC4}
X.~Liu, J.~Sha, H.~Xie, F.~Gao, S.~Jin, Z.~Zhang, X.~You, and C.~Zhang,
  ``Efficient channel estimator with angle-division multiple access,'' {\em
  IEEE Transactions on Circuits and Systems I: Regular Papers}, vol.~66, no.~2,
  pp.~708--718, 2019.

\bibitem{CESoC5}
O.~Castañeda, T.~Goldstein, and C.~Studer, ``Vlsi designs for joint channel
  estimation and data detection in large simo wireless systems,'' {\em IEEE
  Transactions on Circuits and Systems I: Regular Papers}, vol.~65, no.~3,
  pp.~1120--1132, 2018.

\bibitem{zhou2020deep}
R.~Zhou, F.~Liu, and C.~W. Gravelle, ``Deep learning for modulation
  recognition: A survey with a demonstration,'' {\em IEEE Access}, vol.~8,
  pp.~67366--67376, 2020.

\bibitem{9174643}
S.~Zheng, S.~Chen, and X.~Yang, ``Deepreceiver: A deep learning-based
  intelligent receiver for wireless communications in the physical layer,''
  {\em IEEE Transactions on Cognitive Communications and Networking}, vol.~7,
  no.~1, pp.~5--20, 2021.

\bibitem{Himani_DL}
S.~Chandhok, H.~Joshi, A.~V. Subramanyam, and S.~J. Darak, ``Novel deep
  learning framework for wideband spectrum characterization at sub-nyquist
  rate,'' {\em Wireless Networks (Springer)}, vol.~27, pp.~4727--4746, 2021.

\bibitem{wang2018deep}
T.~Wang, C.-K. Wen, S.~Jin, and G.~Y. Li, ``Deep learning-based csi feedback
  approach for time-varying massive mimo channels,'' {\em IEEE Wireless
  Communications Letters}, vol.~8, no.~2, pp.~416--419, 2018.

\bibitem{zhou2019Beammanagement}
P.~Zhou, X.~Fang, X.~Wang, Y.~Long, R.~He, and X.~Han, ``Deep learning-based
  beam management and interference coordination in dense mmwave networks,''
  {\em IEEE Transactions on Vehicular Technology}, vol.~68, no.~1,
  pp.~592--603, 2019.

\bibitem{5200378}
R.~C. Daniels, C.~M. Caramanis, and R.~W. Heath, ``Adaptation in
  convolutionally coded mimo-ofdm wireless systems through supervised learning
  and snr ordering,'' {\em IEEE Transactions on Vehicular Technology}, vol.~59,
  no.~1, pp.~114--126, 2010.

\bibitem{shi2011iteratively}
Q.~Shi, M.~Razaviyayn, Z.-Q. Luo, and C.~He, ``An iteratively weighted mmse
  approach to distributed sum-utility maximization for a mimo interfering
  broadcast channel,'' {\em IEEE Transactions on Signal Processing}, vol.~59,
  no.~9, pp.~4331--4340, 2011.

\bibitem{zhang2021scalable}
X.~Zhang, H.~Zhao, J.~Xiong, X.~Liu, L.~Zhou, and J.~Wei, ``Scalable power
  control/beamforming in heterogeneous wireless networks with graph neural
  networks,'' in {\em 2021 IEEE Global Communications Conference (GLOBECOM)},
  pp.~01--06, IEEE, 2021.

\bibitem{8054694}
T.~O’Shea and J.~Hoydis, ``An introduction to deep learning for the physical
  layer,'' {\em IEEE Transactions on Cognitive Communications and Networking},
  vol.~3, no.~4, pp.~563--575, 2017.

\bibitem{dorner2018Endtoend}
S.~Dörner, S.~Cammerer, J.~Hoydis, and S.~t. Brink, ``Deep learning based
  communication over the air,'' {\em IEEE Journal of Selected Topics in Signal
  Processing}, vol.~12, no.~1, pp.~132--143, 2018.

\bibitem{8052521}
H.~Ye, G.~Y. Li, and B.-H. Juang, ``Power of deep learning for channel
  estimation and signal detection in ofdm systems,'' {\em IEEE Wireless
  Communications Letters}, vol.~7, no.~1, pp.~114--117, 2018.

\bibitem{9739166}
D.~Luan and J.~Thompson, ``Low complexity channel estimation with neural
  network solutions,'' in {\em WSA 2021; 25th International ITG Workshop on
  Smart Antennas}, pp.~1--6, 2021.

\bibitem{9250028}
Y.~Zhang, A.~Doshi, R.~Liston, W.-T. Tan, X.~Zhu, J.~G. Andrews, and R.~W.
  Heath, ``Deepwiphy: Deep learning-based receiver design and dataset for ieee
  802.11ax systems,'' {\em IEEE Transactions on Wireless Communications},
  vol.~20, no.~3, pp.~1596--1611, 2021.

\bibitem{pan2010TransferLearning}
S.~J. Pan and Q.~Yang, ``A survey on transfer learning,'' {\em IEEE
  Transactions on Knowledge and Data Engineering}, vol.~22, no.~10,
  pp.~1345--1359, 2010.

\bibitem{CESoC1}
P.~K. Chundi, X.~Wang, and M.~Seok, ``Channel estimation using deep learning on
  an fpga for 5g millimeter-wave communication systems,'' {\em IEEE
  Transactions on Circuits and Systems I: Regular Papers}, vol.~69, no.~2,
  pp.~908--918, 2022.

\bibitem{CESoC2}
F.~Restuccia and T.~Melodia, ``Deep learning at the physical layer: System
  challenges and applications to 5g and beyond,'' {\em IEEE Communications
  Magazine}, vol.~58, no.~10, pp.~58--64, 2020.

\bibitem{CESoC3}
P.~Sahoo, R.~Rajoria, S.~Chandhok, S.~J. Darak, D.~P. Pau, and H.~Dabral,
  ``Resource constrained neural networks for direction-of-arrival estimation in
  micro-controllers,'' in {\em The First International Conference on
  AI-ML-Systems}, AIMLSystems 2021, (New York, NY, USA), Association for
  Computing Machinery, 2021.

\bibitem{ACM_DLCESD}
S.~Brennsteiner, T.~Arslan, J.~Thompson, and A.~McCormick, ``A real-time deep
  learning ofdm receiver,'' {\em ACM Trans. Reconfigurable Technol. Syst.},
  vol.~15, dec 2022.

\bibitem{Sai_tnnls}
S.~V.~S. Santosh and S.~J. Darak, ``Multiarmed bandit algorithms on zynq
  system-on-chip: Go frequentist or bayesian?,'' {\em IEEE Transactions on
  Neural Networks and Learning Systems}, pp.~1--14, 2022.

\bibitem{sze2017efficient}
V.~Sze, Y.-H. Chen, T.-J. Yang, and J.~S. Emer, ``Efficient processing of deep
  neural networks: A tutorial and survey,'' {\em Proceedings of the IEEE},
  vol.~105, no.~12, pp.~2295--2329, 2017.

\bibitem{ruder2016overview}
S.~Ruder, ``An overview of gradient descent optimization algorithms,'' {\em
  arXiv preprint arXiv:1609.04747}, 2016.

\bibitem{Acosta2007ChModels}
G.~Acosta-Marum and M.~A. Ingram, ``Six time- and frequency- selective
  empirical channel models for vehicular wireless lans,'' {\em IEEE Vehicular
  Technology Magazine}, vol.~2, no.~4, pp.~4--11, 2007.

\end{thebibliography}
\bibliographystyle{ieeetr}
\end{document}